\documentclass{article}

\usepackage{graphics}
\usepackage{graphicx}
\usepackage{amsmath}
\usepackage{amsthm}
\usepackage{amssymb}
\usepackage{amsfonts}
\usepackage{url}
\usepackage{bbold}
\usepackage[utf8]{inputenc}
\usepackage[english]{babel}
\usepackage{float}
\usepackage{listings}
\usepackage[ruled]{algorithm2e}
\usepackage[letterpaper, margin=1in]{geometry}
\usepackage{mathtools}
\usepackage{multirow}
\usepackage{authblk}
\usepackage{natbib}
\usepackage{apalike}
\usepackage{xcolor}
\usepackage{rotating}

\DeclareMathOperator*{\argmax}{arg\,max}

\newtheorem{assumption}{Assumption}

\begin{document}
\title{A Calibration Approach to Transportability and Data-Fusion with Observational Data}
\author[1]{Kevin P. Josey}
\author[2]{Fan Yang}
\author[2]{Debashis Ghosh}
\author[3,4]{Sridharan Raghavan*}

\affil[1]{Department of Biostatistics, Harvard T.H. Chan School of Public Health, Boston, MA}
\affil[2]{Department of Biostatistics and Informatics, Colorado School of Public Health, Aurora, CO}
\affil[3]{Department of Veterans Affairs Eastern Colorado Health Care System, Aurora, CO}
\affil[*]{Corresponding Author: Sridharan Raghavan, sridharan.raghavan@va.gov}

\maketitle

\begin{abstract}

Two important considerations in clinical research studies are proper evaluations of internal and external validity. While randomized clinical trials can overcome several threats to internal validity, they may be prone to poor external validity. Conversely, large prospective observational studies sampled from a broadly generalizable population may be externally valid, yet susceptible to threats to internal validity, particularly confounding. Thus, methods that address confounding and enhance transportability of study results across populations are essential for internally and externally valid causal inference, respectively. These issues persist for another problem closely related to transportability known as data-fusion. We develop a calibration method to generate balancing weights that address confounding and sampling bias, thereby enabling valid estimation of the target population average treatment effect. We compare the calibration approach to two additional doubly-robust methods that estimate the effect of an intervention on an outcome within a second, possibly unrelated target population. The proposed methodologies can be extended to resolve data-fusion problems that seek to evaluate the effects of an intervention using data from two related studies sampled from different populations. A simulation study is conducted to demonstrate the advantages and similarities of the different techniques. We also test the performance of the calibration approach in a motivating real data example comparing whether the effect of biguanides versus sulfonylureas - the two most common oral diabetes medication classes for initial treatment - on all-cause mortality described in a historical cohort applies to a contemporary cohort of US Veterans with diabetes.

\end{abstract}

\section{Introduction}

Two common and related problems in statistics are causal inference and the generalizability of study results to a population of interest. A principal barrier to causal inference is mitigating confounding bias that can afflict the association between an exposure or treatment variable and the outcome. One solution for addressing confounding bias is to conduct a randomized trial. However, a randomized trial is impractical in many scientific and medical contexts. Therefore, methods for causal inference using observational data are essential. Even when valid causal inference can be drawn from a study, the effect estimates may not be accurate for the population of interest. This discordance occurs when the population that is sampled for an observational study or randomized trial - i.e. the \textit{study population} - diverges from the specific population of interest for applying the study results – i.e. the \textit{target population}. For example, the population of patients with a given disease may differ across important characteristics from the population receiving a specific intervention for that disease. In this simplistic scenario, extending valid inferences which evaluate the efficacy of the intervention to the target population containing every patient with the disease is of practical importance. For much of this manuscript, we focus on a more general yet well-defined setup for generalizing results onto a target population known as \textit{transportability} \citep{westreich_transportability_2017}. Whereas generalizability requires the study population to be nested entirely within the target population, transportability allows the study and target populations to be disjoint. More information about the distinction between generalizability and transportability can be found elsewhere \citep{westreich_transportability_2017, josey2021transporting}. We then transition our attention to the problem of \textit{data-fusion} \citep{bareinboim_causal_2016} a closely related problem to transportability, wherein inferences occur on the target population and data for the outcome and treatment assignment are available from two respective samples of the target and study populations. Relative to transportability, where treatment and outcome data are only available for units in the study sample, the data-fusion estimator of a target population causal effect should be more efficient given the greater availability of data.

Methods for causal inference and transporting effect estimates to a target population are two recent topics of substantial interest in the statistical literature. Our goal is to combine methods found in these two respective areas in order to minimize bias due to confounding, which is unavoidable in observational studies, and to account for differences between study and target populations that could influence the causal effect estimate in the population of interest. The propensity score, or the probability of exposure given a set of measured covariates, has emerged as a popular tool in causal inference. Conditioning a causal effect estimator on the propensity score in essence balances the distribution of confounders between the exposed and unexposed participants in an observational study, thereby eliminating confounding bias \citep{rosenbaum_central_1983}. Through a similar mechanism, by conditioning a causal effect estimator on the sampling score, or the probability of being sampled for a study given the measured covariates, we can transport study results onto a target population in the presence of \textit{effect modification} \citep{westreich_transportability_2017}. Extensions and comparisons of methods for transporting observational study results are limited, with most methods opting to focus on transporting results from a randomized controlled trial \citep{rudolph_robust_2017,lee2021improving,dahabreh2020double,josey2021transporting}. Moreover, there is little discussion on extensions for many of these methods to reconcile differences between two observational studies which examine the same exposure/outcome relationship, as in the data-fusion problem \citep{bareinboim_causal_2016}. Further compounding the issues already facing transportability and data-fusion, fitting parametric models of the propensity and sampling scores with maximum likelihood estimation has several limitations, particularly in finite sample settings. One of these issues is model misspecification \cite{kang_demystifying_2007}. Therefore, the proposed implementations for extending inferences, either through transportability or data-fusion, should complement methodological developments found in the causal inference literature that overcome some of these limitations, such as double-robustness to help ameliorate model misspecification.

Calibration estimators \citep{deville_calibration_1992} have had recent success in both transporting causal effects from a randomized controlled trial to a target population and estimating causal effects using observational data. We propose combining the approach of \cite{chan2016globally} and \cite{josey2021framework}, which finds \textit{balancing weights} that correct for confounding bias, with an exponential tilting estimator that estimates \textit{sampling weights} to remove any bias attributable to differences in the covariate distribution between the study participants and non-participants sampled from the target population \citep{signorovitch_comparative_2010}. The exponential tilting approach to estimate sampling weights was used by \cite{lee2021improving} to generalize randomized controlled trial results onto a target population characterized by an observational cohort. Their solution estimates the target population average treatment effect using a class of estimators that augment models for the propensity and sampling scores with a model of the outcome process \citep{robins_estimation_1994}. However, they stop short of identifying calibration weights that will also balance the covariate distribution between the treatment groups. To supplement their results, we propose a full calibration approach for transporting causal effects that produces weights that balance the covariate distribution between the treatment groups and between the study and target samples. We then show how the full calibration approach can be adapted to solve the data-fusion problem, enabling us to combine two observational studies that compare the same exposure and outcome. In the data-fusion setting, we will need to carefully consider additional assumptions appended to those already required for transportability. We demonstrate that the full calibration approach is doubly-robust, meaning that if either the potential outcome model is correctly specified or both the propensity and sampling score models are correctly specified, then the full calibration estimator of the target population average treatment effect is consistent.

To help showcase the proposed calibration solutions, we will evaluate effect estimates for diabetes treatment derived from older observational data on a contemporary real-world patient sample. This scenario is well-suited for transportability and data-fusion as diabetes patient characteristics, which potentially modify the effects of the initial diabetes treatments, have changed substantially over the last 10-20 years \citep{ali_achievement_2013, selvin_trends_2014,  gregg_changes_2014, geiss_prevalence_2014, gregg_trends_2018, cheng_trends_2018, raghavan_diabetes_2019}. The analysis here focuses on the comparative effectiveness of metformin and sulfonylureas - the two most commonly used drugs/classes for initial diabetes treatment in the US \citep{berkowitz_initial_2014, hampp_use_2014, desai_patterns_2012} - as first-line diabetes treatments. Several large observational studies and clinical trials comparing metformin and sulfonylureas have found that sulfonylurea use may be associated with poorer cardiovascular and mortality outcomes than metformin \citep{schramm_mortality_2011, roumie_comparative_2012, wheeler_mortality_2013, hong_effects_2013, varvaki_rados_association_2016}. Consequently, sulfonylurea use has declined over the last decade. However, they remain inexpensive and effective diabetes medications, making them an attractive treatment option in the right patients - for example, those at low risk of treatment related adverse events. With changing patient characteristics and more effective patient selection for sulfonylurea treatment, it is possible that the comparative effectiveness of sulfonylureas relative to metformin in real-world use has improved over time. We apply the transportability and data-fusion methods described in this manuscript to the problem of updating effect estimates of metformin versus sulfonylureas as a first-line diabetes treatment on all-cause mortality in the United States Veterans Affairs (VA) population. Using data available from 2004-2009, we will transport the risk difference of mortality from initial monotherapy with either a sulfonylurea or metformin regimen to a more contemporary cohort of diabetes patients diagnosed between 2010-2014. The estimated effects in the 2010-2014 sample will serve as a benchmark for the 2004-2009 cohort estimates transported to the 2010-2014 population. We also use the data-fusion approach to combine the 2004-2009 and 2010-2014 cohorts to estimate the treatment effect in the 2010-2014 VA diabetic population. Understanding how temporal changes in a health system's patient population can modify the effect estimates from treatment choices for diabetes could have substantial population health impacts.

The remainder of this article is structured as follows. In Section \ref{prelim:04} we introduce the notation and several assumptions that will be referenced throughout. In Section \ref{transport:04}, we describe various methods for transporting causal effects in observational settings, including our proposed full calibration approach. In Section \ref{fusion:04} we propose an extension to the full calibration method in Section \ref{transport:04} so that it is applicable in a data-fusion setting that requires combining observational studies. In Section \ref{simulation:04}, we compare the methods described in Sections \ref{transport:04} and \ref{fusion:04} through simulations. Section \ref{illustrate:04} contains a data analysis using the full calibration approaches to transportability and data-fusion. Finally, we conclude with a discussion in Section \ref{discussion:04}.

\section{Setting and Preliminaries}\label{prelim:04}

\subsection{Notation and Definitions}

The setup for transportability and data-fusion with observational data requires - first and foremost - data from two separate observational studies. Define $S_i\in\{0,1\}$ as a sampling indicator denoting whether the independent sampling unit $i = 1,2,\ldots,n$ is a study non-participant or participant. We sometimes refer to units $\{i:S_i = 1\}$ as the \textit{study sample} and units $\{i:S_i = 0\}$ as the \textit{target sample}. We denote $n_1 = \sum_{i = 1}^n S_i$ and $n_0 = \sum_{i = 1}^n(1 - S_i)$ with $n = n_1+n_0$. We suppose that the non-participants in the target sample represent a random sample from the target population - the population we would like to infer upon. The study sample is representative of the study population which is fundamentally different than the target population.

For each $i = 1,2,\ldots,n$, let $\mathbf{X}_{i}\in\mathcal{X}$ denote a vector of measured covariates, $Y_{i}\in\Re$ denote the outcome and $Z_i\in\{0,1\}$ denote the treatment assignment. We apply the potential outcomes framework \citep{rubin_estimating_1974} to identify the causal estimand of interest and the necessary assumptions for transportability \citep{lesko_generalizing_2017}. Let $Y_{i}(0)$ denote the potential outcome when $Z_i = 0$ and $Y_{i}(1)$ denote the potential outcome when $Z_i = 1$. The conditional expectations of these potential outcomes are denoted $\mu_0(\cdot) \equiv \mathbb{E}[Y_i(0)|\cdot]$ and $\mu_1(\cdot) \equiv \mathbb{E}[Y_i(1)|\cdot]$. The target population average treatment effect is defined as $\tau_0 \equiv \mathbb{E}[Y_i(1)-Y_i(0)|S_i = 0]$. $\tau_0$ is the estimand of interest for both transportability and data-fusion.

Conditioned on $\mathbf{X}_i$, we set $\rho(\mathbf{X}_i) \equiv \Pr\{S_i = 1|\mathbf{X}_i\}$, $\pi_1(\mathbf{X}_i) \equiv \Pr\{Z_i = 1|S_i = 1, \mathbf{X}_i\}$ and $\pi_0(\mathbf{X}_i) \equiv \Pr\{Z_i = 1|S_i = 0,\mathbf{X}_i\}$ for all $i = 1,2,\ldots, n$. Note that the probability of treatment conditioned on the sample indicator and covariates can be alternatively expressed as \[ \pi_s(\mathbf{X}_i) \equiv \Pr\{Z_i = 1|S_i = s, \mathbf{X}_i\} \equiv s\pi_1(\mathbf{X}_i) + (1 - s)\pi_0(\mathbf{X}_i). \] Define $\{c_j(\mathbf{X}): j = 1,2,\ldots, m\}$ as the set of functions that generate linearly independent features to be balanced between treatment groups and the two samples. We will refer to these quantities as \textit{balance functions}. Furthermore, we will assume $c_1(\mathbf{X}_i) = 1$ for all $i = 1,2,\ldots,n$. The target sample moments of the balance functions are defined as $\hat{\theta}_{0j} = n_0^{-1}\sum_{i = 1}^n (1 - S_i) c_j(\mathbf{X}_i)$, which is a consistent and unbiased estimator for $\theta_{0j} \equiv \mathbb{E}[c_j(\mathbf{X}_i)|S_i = 0]$ for all $j = 1,2,\ldots,m$. Finally, we define the logit link function as $g(w) = \log[w/(1 - w)]$ and its inverse $g^{-1}(w) = \exp(w)/[1 + \exp(w)]$.

\subsection{Assumptions for Transportability and Data-Fusion}\label{setup:04}

\subsubsection{Nonparametric Identifiability Assumptions}

Under the potential outcomes model and given the definitions listed in the previous section, we may begin to develop the setting for which transportability and data-fusion methods are applicable using observational data \citep{pearl_external_2014, bareinboim_causal_2016}. We frame the setup to both problems through the following set of assumptions. These assumptions are an extension to those proposed in other articles regarding the transportability of experimental results across populations \citep{rudolph_robust_2017, lee2021improving}. We combined these assumptions with the assumptions necessary for conducting causal inference in the presence of confounding \citep{rubin_estimating_1974}. To begin, we invoke the stable unit treatment value assumption which requires consistency, $Y_i(z) = Y_i$ when $Z_i = z$, and no interference between units \citep{rubin_estimating_1974}. This means the observed outcome is equivalent to $Y_{i} \equiv Z_{i}Y_{i}(1)+(1 - Z_{i})Y_{i}(0)$.
\begin{assumption}[Strongly Ignorable Treatment Assignment]\label{sita:04}
In the case of transportability, the potential outcomes among the study participants are independent of the treatment assignment given $\mathbf{X}_i$, and therefore: $\mathbb{E}[Y_i(z)|S_i = 1, \mathbf{X}_i, Z_i] = \mathbb{E}[Y_i(z)|S_i = 1, \mathbf{X}_i]$ for both $z \in \{0,1\}$. For data-fusion, this condition is required for both study participants and non-participants: $\mathbb{E}[Y_i(z)|S_i, \mathbf{X}_i, Z_i] = \mathbb{E}[Y_i(z)|S_i, \mathbf{X}_i]$ for both $z \in \{0,1\}$.
\end{assumption}
\begin{assumption}[Treatment Effect Exchangeability]\label{exchange:04}
Among all independent sampling units in either the study sample or the target sample, the expected value of the individual treatment effects conditioned on the covariates are exchangeable between samples: $\mathbb{E}[Y_{i}(1) - Y_{i}(0)|S_i, \mathbf{X}_i] = \mathbb{E}[Y_{i}(1) - Y_{i}(0)|\mathbf{X}_i]$ for all $i = 1,2,\ldots,n$.
\end{assumption}
\begin{assumption}[Sample Positivity]\label{positivity-sample:04}
The probability of non-study participation, conditioned on the baseline covariates necessary to ensure treatment effect exchangeability, is bounded below by zero: $\Pr\{S_i = 0|\mathbf{X}_i \} > 0$ for all $i = 1,2,\ldots,n$.
\end{assumption}
\begin{assumption}[Treatment Positivity]\label{positivity-treat:04}
For transportability, the probability of treatment conditioned on the baseline covariates in the study sample is bounded between zero and one: $0 < \Pr\{Z_i = 1|S_i = 1, \mathbf{X}_i\} < 1$ for all $i = 1,2,\ldots,n$. For data-fusion, this condition is required for the both the study and target samples: $0 < \Pr\{Z_i = 1|S_i, \mathbf{X}_i\} < 1$.
\end{assumption}

\subsubsection{Parametric Assumptions for Calibration Estimators}

In addition to Assumptions \ref{sita:04}-\ref{positivity-treat:04}, the following set of assumptions are necessary to establish the doubly-robust properties of the calibration weighted estimates for transportability and data-fusion. For more context, we will show that if either Assumption \ref{linear:04} is satisfied or both Assumptions \ref{odds-sample:04} and \ref{odds-treat:04} hold, then the so-called full calibration method that we present in Section \ref{calibration:04} is consistent. If \ref{odds-sample:04} and \ref{odds-treat:04} hold, then \ref{linear:04} is not required to achieve statistical consistency. Likewise, if Assumption \ref{linear:04} holds then \ref{odds-sample:04} and \ref{odds-treat:04} are not required to ensure statistical consistency.

\begin{assumption}[Conditional Linearity of the Potential Outcomes]\label{linear:04}
The expected value of the potential outcomes, conditioned on $\mathbf{\mathbf{X}}_i$, is linear across the span of the balancing functions while adhering to treatment effect exchangeability (Assumption \ref{exchange:04}): \[ \mu_0(S_i = s, \mathbf{X}_i) = \sum_{j = 1}^m c_j(\mathbf{X}_i)\beta_{sj} \enskip \text{and} \enskip \mu_1(S_i = s, \mathbf{X}_i) = \mu_0(S_i = s, \mathbf{X}_i) +\sum_{j = 1}^m c_j(\mathbf{X}_i)\alpha_{j}\] for all $i = 1,2,\ldots,n$ with $\alpha_j, \beta_{sj}\in\Re$ for all $j = 1,2,\ldots,m$ and $s \in \{0,1\}$.
\end{assumption}
\begin{assumption}[Conditional Linear Log-Odds for Sampling]\label{odds-sample:04}
The log-odds of being in the study sample versus the target sample are linear across the span of the covariates: $g[\rho(\mathbf{X}_i)] = \sum_{j = 1}^m c_j(\mathbf{X}_i)\gamma_j$ for all $i = 1,2,\ldots,n$ and $\gamma_j\in\Re$ for all $j = 1,2,\ldots,m$. 
\end{assumption}
\begin{assumption}[Conditional Linear Log-Odds for Treatment]\label{odds-treat:04}
The probability of treatment in the study sample and the target sample are linear across the span of the covariates: \[ g[\pi_s(\mathbf{X}_i)] = s\sum_{j = 1}^m c_j(\mathbf{X}_i)\lambda_{1j} + (1 - s)\sum_{j = 1}^m c_j(\mathbf{X}_i)\lambda_{0j} \] for all $i = 1,2\ldots,n$ and $\lambda_{0j},\lambda_{1j}\in\Re$ for all $j = 1,2,\ldots,m$ and $S_i = s$. 
\end{assumption}

The linearity conditions found in Assumptions \ref{linear:04}-\ref{odds-treat:04} are relatively strong. Whereas the calibration estimators require the above representations of the potential outcome model, or the sampling and propensity scores, the alternative methods presented in Sections \ref{target:04} and \ref{augment:04} have less stringent requirements to achieve statistical consistency. We will show that the targeted maximum likelihood approach makes the fewest parametric assumptions of all the methods that we will examine whereas the full calibration approach in Section \ref{calibration:04} makes the most parametric assumptions for identification. Having fewer parametric assumptions allows for more intricate modeling choices, including the ability to employ machine learning techniques. However, these methods lose some interpretability provided by the calibration weights, which will exactly balance whichever distributional features are specified in the primal problem between the study and target samples and between the treatment groups - a quality often desired in causal analyses \citep{austin2009balance}. While the TMLE and augmented approaches we present later on do not require parametric assumptions for identification, the methods used to estimate the nuisance parameters can be parametric. If the nuisance parameters are estimated with parametric models, then the TMLE and augmented approaches are also making the same strong parametric assumptions as the calibration approach. We demonstrate this fact in our simulation study in order to evaluate the doubly-robust properties of the various estimators outlined for $\tau_0$.

\subsubsection{Data-Fusion Specific Assumptions}

The distinction between transportability and data-fusion essentially amounts to how much data we are provided from the study and target samples. For problems of transportability, we require the complete individual-level data from study sample, but only the individual-level covariate data from the target sample (i.e. $\mathbf{X}_i$ for all $\{i:S_i = 0\}$). In data-fusion, both the study and target samples provide data on $\mathbf{X}_i$, $Y_i$, and $Z_i$ for all $i = 1,2,\ldots,n$. It should not be a surprise that the latter setting is more powerful given the additional data. However, in many data analysis applications, $Y_i$ and $Z_i$ are not available from the target sample leaving data-fusion impracticable. In those cases, transportability is an appealing alternative.

\begin{assumption}[Potential Outcome Exchangeability]\label{out-exchange:04}
Among all independent sampling units in either the study or target sample, the expected value of the potential outcomes conditioned on the covariates are exchangeable between samples: $\mathbb{E}[Y_i(z)|S_i, \mathbf{X}_i] = \mathbb{E}[Y_i(z)|\mathbf{X}_i]$ for both $z \in \{0,1\}$. 
\end{assumption}
\begin{assumption}[Propensity Score Exchangeability]\label{ps-exchange:04}
For all $\mathbf{X}_i \in \mathcal{X}$, we have $\pi_1(\mathbf{X}_i) = \pi_0(\mathbf{X}_i)$.
\end{assumption}

The exchangeability assumptions of Assumptions \ref{out-exchange:04} and \ref{ps-exchange:04} are key when considering the model design for estimators of $\tau_0$ in the data-fusion setting. Potential outcome exchangeability is a stronger assumption than the assumptions of treatment effect exchangeability (Assumption \ref{exchange:04}). The former requires the expected values of the potential outcomes be exchangeable, conditional on the effect modifiers, confounders, and prognostic variables. We refer to \textit{prognostic variables} as covariate measurements that are predictive of the outcome, but are otherwise uncorrelated with the treatment to be exclusive of confounders. The treatment effect exchangeability assumption, on the other hand, requires exchangeability conditional on both the effect modifiers to eliminate sampling bias and the confounders to eliminate confounding bias. Potential outcome exchangeability is a key limitation to more general approaches that impute potential outcomes for specific populations without necessarily evaluating a treatment effect \citep{keele2020hospital, hirshberg2019minimax}. Since comparable methods like the augmented approach \citep{lee2021improving} and the target maximum likelihood approach \citep{rudolph_robust_2017} that we will introduce in the next section require estimates of $\mu_0(S_i, \mathbf{X}_i)$ and $\mu_1(S_i, \mathbf{X}_i)$, potential outcome exchangeability becomes important if not necessary whenever both the study and target samples are used to estimate $\tau_0$. Propensity score exchangeability assumes the propensity score is the same across both populations. This assumption is not an issue for transporting observational study results since these problems implicitly assume that the treatment assignment mechanism is the same for both samples. This is not an unreasonable assumption given that $Z_i$ is not observed in the target sample in the transportability case. In the data-fusion setting, where the distribution of the treatment assignment can vary between samples, propensity score exchangeability becomes an important consideration. If both Assumptions \ref{out-exchange:04} and \ref{ps-exchange:04} are violated, then $S_i$ behaves like a surrogate confounder \citep{vanderweele2013definition} and omitting it from either the potential outcome models or the propensity score model may lead to biased effect estimates.

\section{Methods for Transporting Observational Results}\label{transport:04}

\subsection{Targeted Maximum Likelihood}\label{target:04}

Targeted maximum likelihood estimation (TMLE) has emerged as a flexible framework for estimating a variety of causal estimands \citep{laan_targeted_2006}. Specifically \cite{rudolph_robust_2017} apply this framework to estimate $\tau_0$ within the transportability setting described in Section \ref{setup:04}. TMLE is adapted from the G-computation family of estimators which finds $\hat{\mu}_0(S_i = 1, \mathbf{X}_i) \equiv \hat{\mathbb{E}}(Y_i| S_i = 1, \mathbf{X}_i, Z_i = 0)$ and $\hat{\mu}_1(S_i = 1, \mathbf{X}_i) \equiv \hat{\mathbb{E}}(Y_i| S_i = 1, \mathbf{X}_i, Z_i = 1)$ using the independent sampling units $\{i:S_i = 1\}$ to solve for 
\begin{equation}\label{gcomp:04}
\hat{\tau}_\text{G} = \frac{1}{n_0} \sum_{\{i:S_i = 0\}} \left[\hat{\mu}_1(S_i = 1, \mathbf{X}_i) - \hat{\mu}_0(S_i = 1, \mathbf{X}_i) \right].
\end{equation} For this estimator, if we can show $\hat{\mu}_1( S_i = 1, \mathbf{X}_i) \rightarrow_p \mu_1(S_i = 1, \mathbf{X}_i)$ and $\hat{\mu}_0(S_i = 1, \mathbf{X}_i) \rightarrow_p \mu_0(S_i = 1, \mathbf{X}_i)$, then given Assumptions \ref{sita:04}-\ref{exchange:04} we have $\hat{\tau}_G \rightarrow_p \tau_0$.

\cite{rudolph_robust_2017} extend this intuitive approach to account for potential bias induced from misspecifying $\hat{\mu}_0(S_i = 1, \mathbf{X}_i)$ and $\hat{\mu}_1(S_i = 1, \mathbf{X}_i)$. First, the proposed TMLE update will find estimates of the expected potential outcomes on the logit scale \citep{gruber2010targeted}. This requires standardizing $Y_i^* \equiv \left[Y_i - Y^{-}\right]\left[Y^{+} - Y^{-}\right]^{-1}$ where $Y^{-} \equiv \min_{\{i:S_i = 1\}}(Y_i)$ and $Y^{+} \equiv \max_{\{i:S_i = 1\}}(Y_i)$ from the available outcome data. Additionally, we will also need to find $\hat{\mu}^*_0(S_i = 1, \mathbf{X}_i) \equiv \hat{\mathbb{E}}[Y_i^*|S_i = 1, \mathbf{X}_i, Z_i = 0]$ and $\hat{\mu}^*_1(S_i = 1, \mathbf{X}_i) \equiv \hat{\mathbb{E}}[Y_i^*| S_i = 1, \mathbf{X}_i, Z_i = 1]$. The TMLE solution updates the estimates of the conditional means for the transformed potential outcomes using consistent estimates for $\rho(\mathbf{X}_i)$ and $\pi_1(\mathbf{X}_i)$, which we denote as $\hat{\rho}(\mathbf{X}_i) = \hat{\Pr}\left\{S_i = 1|\mathbf{X}_i\right\}$ and $\hat{\pi}_1(\mathbf{X}_i) \equiv \hat{\Pr}\left\{Z_i = 1|S_i = 1, \mathbf{X}_i\right\}$. These estimators are combined into a so-called clever covariate \citep{schuler_targeted_2017} to find
\[ \begin{split}
\hat{\eta}^*_0( S_i = 1, \mathbf{X}_i) &= g^{-1}\left\{g\left[\hat{\mu}^*_0(S_i = 1, \mathbf{X}_i)\right] + \hat{\epsilon}_0  \frac{\left[1 - \hat{\rho}(\mathbf{X}_i)\right]}{\hat{\rho}(\mathbf{X}_i)\left[1 - \hat{\pi}_1(\mathbf{X}_i)\right]}\right\} \enskip \text{and}\\
\hat{\eta}^*_1(S_i = 1, \mathbf{X}_i) &= g^{-1}\left\{g\left[\hat{\mu}^*_1(S_i = 1, \mathbf{X}_i)\right] + \hat{\epsilon}_1  \frac{\left[1 - \hat{\rho}(\mathbf{X}_i)\right]}{\hat{\rho}(\mathbf{X}_i)\hat{\pi}_1(\mathbf{X}_i)}\right\}.
\end{split}\]

Estimates of the fluctuation coefficients $\hat{\boldsymbol{\epsilon}} \equiv (\hat{\epsilon}_0, \hat{\epsilon}_1)^T$ are found by projecting the transformed outcomes $Y_i^*$ onto the clever covariates 
\[\hat{h}_0(\mathbf{X}_i) = \frac{(1 - Z_i)[1 - \hat{\rho}(\mathbf{X}_i)]}{\hat{\rho}(\mathbf{X}_i)[1 - \hat{\pi}_1(\mathbf{X}_i)]} \enskip \text{and} \enskip \hat{h}_1(\mathbf{X}_i) = \frac{Z_i[1 - \hat{\rho}(\mathbf{X}_i)]}{\hat{\rho}(\mathbf{X}_i)\hat{\pi}_1(\mathbf{X}_i)},\] 
with $\hat{\mathbf{h}}(\mathbf{X}_i) \equiv \left[\hat{h}_0(\mathbf{X}_i), \hat{h}_1(\mathbf{X}_i)\right]^T$. In this regression, the initial estimates of the potential outcome means $\hat{\mu}^*_0(S_i = 1,\mathbf{X}_i)$ and $\hat{\mu}^*_1(S_i = 1, \mathbf{X}_i)$ serve as offsets. Analytically, estimates of the fluctuation coefficients $\hat{\boldsymbol{\epsilon}}$ are found such that
\begin{equation}\label{fluctuation:04}
\sum_{\{i:S_i = 1\}} \hat{\mathbf{h}}(\mathbf{X}_i) \left(Y_i^* - g^{-1}\left\{Z_i g\left[\hat{\mu}^*_1(S_i = 1,\mathbf{X}_i)\right] + (1 - Z_i)g\left[\hat{\mu}^*_0(S_i = 1,\mathbf{X}_i)\right] + \hat{\mathbf{h}}^T(\mathbf{X}_i)\hat{\boldsymbol{\epsilon}}\right\}\right) = 0.
\end{equation}
Since this regression requires $Y^*_i$ and $Z_i$ to be observed, estimates are calculated using the units $\{i:S_i = 1\}$. The TMLE estimate of $\tau_0$ has a similar form to the G-computation setup, solving for 
\begin{equation}\label{tmle:04}
\hat{\tau}_{\text{TMLE}} = \frac{1}{n_0} \sum_{\{i:S_i = 0\}} \left[\hat{\eta}_1(S_i = 1, \mathbf{X}_i) - \hat{\eta}_0(S_i = 1, \mathbf{X}_i)\right] 
\end{equation}
where $\hat{\eta}_0(S_i = 1, \mathbf{X}_i) = \left[Y^{+} - Y^{-}\right]\hat{\eta}^*_0(S_i = 1,\mathbf{X}_i) + Y^{-}$ and $\hat{\eta}_1(S_i = 1,\mathbf{X}_i) = \left[Y^{+} - Y^{-}\right]\hat{\eta}^*_1(S_i = 1,\mathbf{X}_i) + Y^{-}$. Given Assumptions \ref{sita:04}-\ref{positivity-treat:04}, \cite{rudolph_robust_2017} show that the TMLE estimator for $\tau_0$ is consistent if either the potential outcome models are consistent, or if $\hat{\rho}(\mathbf{X}_i) \rightarrow_p \rho(\mathbf{X}_i)$ and $\hat{\pi}_1(\mathbf{X}_i) \rightarrow_p \pi_1(\mathbf{X}_i)$. \cite{gruber2010targeted} argue that using the logit scale within TMLE constrains the fluctuation coefficients such that $Y^{-} \le \hat{\eta}_0(S_i = 1,\mathbf{X}_i) \le Y^{+}$ and $Y^{-} \le \hat{\eta}_1(S_i = 1,\mathbf{X}_i) \le Y^{+}$, thus avoiding extrapolations of the potential outcome mean estimates while retaining the doubly-robust properties inherent with TMLE.

\subsection{Potential Outcome Model Augmentation}\label{augment:04}

Another doubly-robust method for transporting causal effects is to augment the potential outcome model with weights derived from estimates of the propensity and sampling scores, similar to the implementation of the clever covariates used in TMLE \citep{lee2021improving}. This method also builds upon the intuition surrounding the G-computation estimator found in (\ref{gcomp:04}). Below we present a brief summary of the augmented approach relevant for the transportability setting as opposed to the generalizability setting \citep{josey2021transporting}. For a more complete description about how the method is applied in the generalizability setting, we suggest referring to the original text of \cite{lee2021improving}.

The augmented approach proceeds by fitting component models of the data generating mechanisms for $S_i$, $Y_i$, and $Z_i$ given Assumptions \ref{sita:04}-\ref{positivity-treat:04}. While several estimators for the other nuisance parameters will suffice, the inverse odds of sampling are specifically estimated to satisfy the primal constrained optimization problem to
\begin{equation}\label{primal-eb:04}
\begin{split} 
\text{minimize} &\enskip \sum_{\{i:S_i = 1\}} \left\{q(S_i,\mathbf{X}_i) \log\left[q(S_i,\mathbf{X}_i)\right] - q(S_i,\mathbf{X}_i) \right\} \\
\text{subject to} &\enskip \sum_{i = 1}^n  S_i q(S_i,\mathbf{X}_i)c_j(\mathbf{X}_i) = n_1 \hat{\theta}_{0j}
\end{split}
\end{equation}
for all $j = 1,2,\ldots,m$.Equation (\ref{primal-eb:04}) can equivalently be solved with a Lagrangian dual problem which finds
\begin{equation}\label{dual-eb:04}
\hat{\boldsymbol{\gamma}} = \argmax_{\boldsymbol{\gamma} \in \Re^{m}} \ \sum_{\{i:S_i = 1\}} \left\{ -\exp\left[-S_i\sum_{j = 1}^m  c_j(\mathbf{X}_i) \gamma_j\right] - \sum_{j = 1}^m \hat{\theta}_{0j} \gamma_j \right\}
\end{equation}
where $\hat{\boldsymbol{\gamma}} \equiv (\hat{\gamma}_1, \hat{\gamma}_2, \ldots, \hat{\gamma}_m)^T$. A Lagrangian dual is an unconstrained optimization problem derived by applying the Lagrangian multiplier theorem to a primal constrained convex optimization problem, like the one in (\ref{primal-eb:04}). With the dual solution to (\ref{dual-eb:04}), the resulting balancing weights that satisfy (\ref{primal-eb:04}) can be found with
\begin{equation}\label{weights-eb:04}
\hat{q}(S_i,\mathbf{X}_i) = \exp\left[-S_i\sum_{j = 1}^m c_j(\mathbf{X}_i)\hat{\gamma}_j \right].
\end{equation}
With the estimated sampling weights, the conditional mean estimates of the potential outcomes, and an appropriate propensity score model, \cite{lee2021improving} construct an augmented estimator for $\tau_0$ which solves for
\begin{equation}\label{aug:04}
\begin{split}
\hat{\tau}_{\text{AUG}} &= \frac{1}{n_1} \sum_{\{i:S_i = 1\}}\hat{q}(S_i,\mathbf{X}_i)\left[\frac{Z_i[Y_i - \hat{\mu}_1(S_i = 1, \mathbf{X}_i)]}{\hat{\pi}_1(\mathbf{X}_i)} - \frac{(1 - Z_i)[Y_i - \hat{\mu}_0(S_i = 1,\mathbf{X}_i)]}{1 - \hat{\pi}_1(\mathbf{X}_i)}\right] \\ &\qquad + \frac{1}{n_0} \sum_{\{i:S_i = 0\}} \left[\hat{\mu}_1(S_i = 1,\mathbf{X}_i) - \hat{\mu}_0(S_i = 1,\mathbf{X}_i)\right]. 
\end{split}
\end{equation} 

The estimated inverse odds of sampling weights have the property that $\sum_{i = 1}^n S_i \hat{q}(S_i,\mathbf{X}_i) c_j(\mathbf{X}_i) = \sum_{i = 1}^n (1 - S_i)c_j(\mathbf{X}_i)$ for all $j = 1,2,\ldots,m$. In other words, the weighted sample moments of the balance functions in the study sample are equal to the unweighted sample moments of the balance functions in the target sample. Under Assumptions \ref{sita:04}-\ref{positivity-treat:04}, the augmented estimator $\hat{\tau}_{\text{AUG}}$ can be shown to be doubly-robust \citep{lee2021improving}. We can easily see this result given the following heuristic. In the scenario where $\hat{\mu}_0(S_i = 1, \mathbf{X}_i) \rightarrow_p \mu_0(S_i = 1,\mathbf{X}_i)$, $\hat{\mu}_1(S_i = 1,\mathbf{X}_i) \rightarrow_p \mu_1(S_i = 1,\mathbf{X}_i)$, the first summation in (\ref{aug:04}) has an expected value of zero while the second summation is consistent for $\tau_0$. When Assumption \ref{odds-sample:04} holds, if $\hat{\pi}_1(\mathbf{X}_i) \rightarrow_p \pi_1(\mathbf{X}_i)$, then the first sum in (\ref{aug:04}) will consistently negate the bias produced by the second summation.

\subsection{A Full Calibration Approach to Transportability}\label{calibration:04}

Our proposed solution to the problem of transporting observational study results combines the calibration weighting approach for correcting confounding bias encountered within causal effect estimation \citep{josey2021framework} with the calibrated sampling weights used in Section \ref{augment:04}, which removes bias induced by the differences of the covariate distribution between the two samples \citep{signorovitch_comparative_2010,lee2021improving}. In other words, we estimate a vector of weights that concurrently balance both the samples and treatment groups.

The combined balancing and sampling weights are estimated contemporaneously in a manner similar to (\ref{dual-eb:04}) and (\ref{weights-eb:04}). The full calibration weights that balance both the treatment and sampling groups are estimated with the primal problem to
\begin{equation}\label{primal-cal:04}
\begin{split} 
\text{minimize} &\enskip \sum_{\{i:S_i = 1\}} \left\{p(S_i, \mathbf{X}_i, Z_i) \log\left[p(S_i, \mathbf{X}_i, Z_i)\right] - p(S_i, \mathbf{X}_i, Z_i) \right\} \\
\text{subject to} &\enskip \sum_{i = 1}^n S_i (2Z_i - 1) p(S_i, \mathbf{X}_i, Z_i)c_j(\mathbf{X}_i) = 0 \enskip \text{(Propensity Score) and} \\
&\enskip \sum_{i = 1}^n  S_i p(S_i, \mathbf{X}_i, Z_i) c_j(\mathbf{X}_i) = n_1 \hat{\theta}_{0j} \enskip \text{(Sampling Score)}
\end{split}
\end{equation}
for all $j = 1,2,\ldots,m$. Much like with the primal-dual relationship between (\ref{primal-eb:04}) and (\ref{dual-eb:04}), the primal problem in (\ref{primal-cal:04}) can be equivalently solved by finding
\begin{equation}\label{dual-cal:04}
\left(\hat{\boldsymbol{\lambda}}^T, \hat{\boldsymbol{\gamma}}^T\right)^T = \argmax_{(\boldsymbol{\lambda}^T, \boldsymbol{\gamma}^T)^T \in \Re^{2m}} \ \sum_{\{i:S_i = 1\}} \left\{ - \exp\left[-(2Z_i - 1) \sum_{j = 1}^m  c_j(\mathbf{X}_i)\lambda_{j} - \sum_{j = 1}^m  c_j(\mathbf{X}_i)\gamma_{j} \right] - \sum_{j = 1}^m \hat{\theta}_{0j}\gamma_{j}\right\}
\end{equation} 
where $\hat{\boldsymbol{\lambda}} \equiv (\hat{\lambda}_{1}, \hat{\lambda}_{2}, \ldots, \hat{\lambda}_{m})^T$ which are then used to generate the balancing weights that satisfy (\ref{primal-cal:04}) with 
\begin{equation}\label{weights-cal:04}
\hat{p}(S_i,\mathbf{X}_i,Z_i) = \exp\left[-S_i(2Z_i - 1)\sum_{j = 1}^m  c_j(\mathbf{X}_i)\hat{\lambda}_{j} - S_i \sum_{j = 1}^m  c_j(\mathbf{X}_i)\hat{\gamma}_{j} \right].
\end{equation} 
The desired estimand $\tau_0$ is estimated using the Hajek-type estimator
\begin{equation}\label{hajek:04}
\hat{\tau}_{\text{CAL}} = \frac{1}{n_1}\sum_{\{i: S_i = 1\}} \hat{p}(S_i, \mathbf{X}_i, Z_i)(2Z_i - 1)Y_i.
\end{equation}

The balancing weights from (\ref{weights-cal:04}) achieve exact balance between the treatment-specific sample moments of the balance functions in the study sample and balances the balance function sample moments between the study and target samples. By rearranging the constraints of (\ref{primal-cal:04}) we get \[ \sum_{i = 1}^n S_i Z_i\hat{p}(S_i, \mathbf{X}_i, Z_i)c_j(\mathbf{X}_i) = \sum_{i = 1}^n S_i(1-Z_i)\hat{p}(S_i,\mathbf{X}_i,Z_i)c_j(\mathbf{X}_i) = \frac{n_1\hat{\theta}_{0j}}{2} .\] Consequently, the estimator in (\ref{dual-cal:04})-(\ref{hajek:04}) is equivalent to the estimator proposed in \cite{josey2021transporting} which was intended for transporting randomized experiments. Combining the results of the sampling and balancing weights allows us to apply M-estimation techniques to show that the double-robustness property holds under Assumption \ref{linear:04} or under the combination of Assumptions \ref{odds-sample:04} and \ref{odds-treat:04} \citep{stefanski_calculus_2002}. Additional details regarding modes of inference about $\tau_0$ with $\hat{\tau}_{\text{CAL}}$ can be found in the online supplement to \cite{josey2021transporting}.

\section{A Full Calibration Approach for Data-Fusion}\label{fusion:04}

There has been limited discussion in the literature on how the methods for transportability presented in Section \ref{transport:04} might be adapted for data-fusion. Recall that for the data-fusion setting, we are provided $\mathbf{X}_i$, $Y_i$, and $Z_i$ for all $i = 1,\ldots,n$. The suggestion by \cite{lee2021improving} is to use the outcome and treatment data from the target sample to find estimates $\tilde{\mu}_0(S_i = 0, \mathbf{X}_i) \equiv \hat{\mathbb{E}}[Y_i|S_i = 0, \mathbf{X}_i, Z_i = 0]$ and $\tilde{\mu}_1(S_i = 0, \mathbf{X}_i) \equiv \hat{\mathbb{E}}[Y_i|S_i = 0, \mathbf{X}_i, Z_i = 1]$, but otherwise keep the procedure in Section \ref{augment:04} the same. Invoking potential outcome exchangeability, we can find estimates of $\tau_0$ by solving for
\begin{equation}\label{aug-fusion:04}
\begin{split}
    \tilde{\tau}_{\text{AUG}} &= \frac{1}{n_1} \sum_{\{i:S_i = 1\}}\hat{q}(S_i,\mathbf{X}_i)\bigg\{\frac{Z_i[Y_i - \tilde{\mu}_1(S_i = 0, \mathbf{X}_i)]}{\hat{\pi}_1(\mathbf{X}_i)} - \frac{(1 - Z_i)[Y_i - \tilde{\mu}_0(S_i = 0, \mathbf{X}_i)]}{1 - \hat{\pi}_1(\mathbf{X}_i)}\bigg\} \\
    &\qquad+ \frac{1}{n_0} \sum_{\{i:S_i = 0\}} \left[\tilde{\mu}_1(S_i = 0, \mathbf{X}_i) - \tilde{\mu}_0(S_i = 0, \mathbf{X}_i)\right]
\end{split}
\end{equation}

This simple modification to (\ref{aug:04}) implies that only units $\{i:S_i = 1\}$ are used to de-bias estimates of the potential outcome means, assuming the potential outcome models are biased. The reasoning for not using all $i = 1,2,\ldots,n$ for the first line of (\ref{aug-fusion:04}) is due to potential conflicts with propensity score exchangeability. Assuming potential outcome exchangeability holds but propensity score exchangeability is violated, then in order to use all available data to de-bias the second line of (\ref{aug-fusion:04}) requires estimates of both $\pi_0(\mathbf{X}_i)$ in addition to $\pi_1(\mathbf{X}_i)$. This is because $\hat{\pi}_1(\mathbf{X}_i)$ would produce biased predictions from an incorrectly specified model of $\pi_0(\mathbf{X}_i)$. For now, the estimator in (\ref{aug-fusion:04}) does not require this assumption at the cost of losing a considerable degree of efficiency that a data-fusion estimator should otherwise gain. Even if specific models for $\pi_1(\mathbf{X}_i)$ and $\pi_0(\mathbf{X}_i)$ are fit in cases where propensity score exchangeability is violated, (\ref{aug-fusion:04}) is still subject to potential outcome exchangeability, unlike its transportability analogue. Let's suppose that the outcome model is correctly specified for the target sample. If potential outcome exchangeability is violated, then the first line of (\ref{aug-fusion:04}) would likely introduce unchecked bias to $\tilde{\tau}_{\text{AUG}}$, even when the propensity score model of the study sample is correctly specified. This assumption could be counteracted in a similar way to counteracting violations to propensity score exchangeability -- by fitting separate outcome models for the two samples and substituting them into (\ref{aug-fusion:04}) accordingly. Because of the extended exchangeability assumptions (Assumptions \ref{out-exchange:04} and \ref{ps-exchange:04}), the design decisions for data-fusion are more complex than the designs for transportability estimators.

We are intentional in constructing the full calibration data-fusion estimator such that it will not require either Assumptions \ref{out-exchange:04} or \ref{ps-exchange:04} while still using all available data. To do this, ultimately, the full calibration approach to data-fusion will need to satisfy two additional constraints to balance treatment groups across both samples relative to the transportability approach in Section \ref{calibration:04}. This objective is not achieved by the primal problem in (\ref{primal-cal:04}). The new goal is to estimate a set of dual vectors which can generate weights to balance the treatment groups while retaining the doubly-robust property. Once we estimate the new calibration weights, we can simply change the index of the summation in (\ref{hajek:04}) to estimate $\tau_0$ over all $i = 1,2,\ldots,n$.

For data-fusion, the primal problem of (\ref{primal-cal:04}) is expanded upon to become
\begin{equation}\label{primal-fusion:04}
\begin{split} 
\text{minimize} &\enskip \sum_{i = 1}^n \left\{p(S_i, \mathbf{X}_i, Z_i) \log\left[p(S_i, \mathbf{X}_i, Z_i)\right] - p(S_i, \mathbf{X}_i, Z_i) \right\} \\
\text{subject to} &\enskip \sum_{i = 1}^n  S_i(2Z_i - 1) p(S_i, \mathbf{X}_i, Z_i) c_j(\mathbf{X}_i) = 0 \enskip 
\text{(Study Sample Propensity Score)}, \\
&\enskip \sum_{i = 1}^n  S_i p(S_i, \mathbf{X}_i, Z_i) c_j(\mathbf{X}_i) = n_1\hat{\theta}_{0j} \enskip \text{(Sampling Score)}, \\
&\enskip \sum_{i = 1}^n  (1 - S_i)(2Z_i - 1) p(S_i, \mathbf{X}_i, Z_i) c_j(\mathbf{X}_i) = 0 \enskip \text{(Target Sample Propensity Score) and} \\
&\enskip \sum_{i = 1}^n  (1 - S_i) p(S_i, \mathbf{X}_i, Z_i) c_j(\mathbf{X}_i) = n_0 \hat{\theta}_{0j} \enskip \text{(Weight Stabilization)}
\end{split}
\end{equation}
\text{for all} \enskip j = 1,2,\ldots,m. Balancing the treatment specific moments in the study sample alone is not sufficient to also balance the treatment groups in the target sample that is required for data-fusion. For this reason, we added two extra constraints to the primal problem of (\ref{primal-cal:04}) resulting in (\ref{primal-fusion:04}). Only the first three constraints would be required if we knew how to correctly specify the sampling and propensity scores, following Assumptions \ref{odds-sample:04} and \ref{odds-treat:04}. The fourth constraint corrects the sampling variation from the target sample and is needed to ensure double-robustness when Assumption \ref{linear:04} holds. This constraint stabilizes the weights in the same way that the \textit{improved} covariate balance propensity score formulation, thereby making the original proposal for the covariate balance propensity score estimator doubly-robust \citep{fan_improving_2018}. These constraints may be relaxed and altered further when the potential outcome exchangeability or the propensity score exchangeability assumptions are fulfilled. Careful attention should also be given to the size of the influence attributable to the respective samples after weighting. Note that the sum of the weights on the right hand side of the second and fourth equality constraints can be modified if more weight should be assigned to one sample versus the other.

Despite the changes to (\ref{primal-cal:04}) expressed in (\ref{primal-fusion:04}), the primal-dual relationship is unaffected by the added constraints, barring any collinear balance functions. A new dual objective can be derived in the same manner as in Section \ref{calibration:04} to find
\begin{equation}\label{dual-fusion:04}
\left(\tilde{\boldsymbol{\lambda}}_s^T, \tilde{\boldsymbol{\gamma}}_s^T\right)^T = \argmax_{(\boldsymbol{\lambda}^T, \boldsymbol{\gamma}^T)^T \in \Re^{2m}} \ \sum_{\{i:S_i = s\}} \left\{ - \exp\left[-(2Z_i - 1) \sum_{j = 1}^m  c_j(\mathbf{X}_i)\lambda_{j} - \sum_{j = 1}^m  c_j(\mathbf{X}_i)\gamma_{j} \right] - \sum_{j = 1}^m \hat{\theta}_{0j}\gamma_{j}\right\}
\end{equation} 
where $\tilde{\boldsymbol{\lambda}}_{s} \equiv (\tilde{\lambda}_{s1}, \tilde{\lambda}_{s2}, \ldots, \tilde{\lambda}_{sm})^T$ and $\tilde{\boldsymbol{\gamma}}_{s} \equiv (\tilde{\gamma}_{s1}, \tilde{\gamma}_{s2}, \ldots, \tilde{\gamma}_{sm})^T$ for $s \in \{0,1\}$. The balancing weights combine the four estimated dual vectors to produce a solution to (\ref{primal-fusion:04}) with 
\begin{equation}\label{weights-fusion:04}
\begin{split}
\tilde{p}(S_i, \mathbf{X}_i, Z_i) &= \exp\Bigg\{-S_i\left[(2Z_i - 1)\sum_{j = 1}^m c_j(\mathbf{X}_i)\tilde{\lambda}_{1j} + \sum_{j = 1}^m c_j(\mathbf{X}_i)\tilde{\gamma}_{1j}\right] \\
&\qquad - (1 - S_i)\left[(2Z_i - 1)\sum_{j = 1}^m c_j(\mathbf{X}_i)\tilde{\lambda}_{0j} + \sum_{j = 1}^m c_j(\mathbf{X}_i)\tilde{\gamma}_{0j}\right]\Bigg\}.
\end{split}
\end{equation} 
The updated Hajek-estimator finds estimates of $\tau_0$ by solving for
\begin{equation}\label{hajek-fusion:04}
\tilde{\tau}_{\text{CAL}} = \frac{1}{n}\sum_{i = 1}^n \tilde{p}(S_i,\mathbf{X}_i,Z_i)(2Z_i - 1)Y_i.
\end{equation}
The summation in (\ref{hajek-fusion:04}) uses all available data, $i = 1,2,\ldots,n$, meaning this estimator should be more efficient than the transportability proposal in (\ref{hajek:04}). Neither the potential outcome exchangeability assumption nor the propensity score exchangeability assumption are necessarily required for the above approach, but they can be helpful to simplify some of the constraints in (\ref{primal-fusion:04}). This might be important in small sample settings where feasible solutions to (\ref{primal-fusion:04}) are difficult or impossible to achieve. For more information on how inference is conducted using this estimator, please refer to the online supplemental material.

\section{Simulation Study}\label{simulation:04}

\subsection{Simulation Setup}\label{sim-setup:04}

To demonstrate the efficacy of the different methods for transporting observational results, and to demonstrate the advantages of using data-fusion when it is feasible, we will conduct a simulation study that evaluates the performance of the three doubly-robust methods for transporting observational results that we have identified in Section \ref{transport:04} and the two data-fusion methods in Section \ref{fusion:04}. We will examine the performance of the proposed methodologies across a range of scenarios that test these methods when violations to the assumptions in Section \ref{setup:04} occur. To establish the doubly-robust quality of these estimators, we must consider possibilities where either the sampling and propensity score models are misspecified or the potential outcome models are misspecified. We also consider the sensitivity of these methods to practical violations to overlap. Finally, we examine the implications of Assumptions \ref{out-exchange:04} and \ref{ps-exchange:04}, which are particularly important for data-fusion. When potential outcome exchangeability holds but propensity score exchangeability does not, then $S_i$ can be viewed as an \textit{instrumental} variable as the sample indicator would affect the treatment assignment but would not have a direct impact on the outcome process that is not through the treatment assignment. Conversely, when propensity score exchangeability holds and potential outcome exchangeability is violated, then $S_i$ is regarded as a nonconfounding \textit{prognostic} variable in the sense that it predicts the outcome but does not confound the relationship between the treatment assignment and the outcome.

The scenarios we examine vary the sample size $n \in \{500,1{,}000,2{,}000\}$, the generative process that determines the treatment assignment through $\pi_s(\mathbf{X}_i)$, the potential outcome processes through $\mu_0(S_i, \mathbf{X}_i)$ and $\mu_1(S_i, \mathbf{X}_i)$, and the sampling mechanism $\rho(\mathbf{X}_i)$. These scenarios are detailed in Table \ref{generative:04}. The covariate data $\mathbf{X}_i \equiv (X_{i1}, X_{i2}, X_{i3}, X_{i4})^{T}$ are distributed as $X_{i1},X_{i2},X_{i3},X_{i4} \sim \mathcal{N}(0, 1)$ for every $i = 1,2,\ldots,n$. We also generate the transformed variables $U_{i1} = \exp\left[(X_{i1} + X_{i4})/2\right]$, $U_{i2} = X_{i2}/\left[1 + \exp(X_{i1})\right]$, $U_{i3} = \log\left(|X_{i2}X_{i3}|\right)$, and $U_{i4} = \left(X_{i3} + X_{i4}\right)^2$ which are used in instances of model misspecification. Each entry in the vector $\mathbf{U}_i \equiv (U_{i1}, U_{i2}, U_{i3}, U_{i4})^{T}$ is standardized to have a mean of zero and marginal variances of one - the same as for $\mathbf{X}_i$. The sample indicators are generated assuming $S_i \sim \text{Bern}\left[\rho(\mathbf{X}_i)\right]$. Treatment assignments are generated by sampling from $Z_{i} \sim \text{Bern}\left[\pi_{S_i}(\mathbf{X}_i)\right]$. Finally, we generate the potential outcomes as $Y_i(0) \sim \mathcal{N}[\mu_0(\mathbf{X}_i. S_i), 4]$ and $Y_i(1) \sim \mathcal{N}[\mu_1(\mathbf{X}_i, S_i), 4]$, which we use to generate the observed outcome $Y_i = Z_i Y_i(1) + (1 - Z_i)Y_i(0)$.

\begin{table}[H]
\tiny
\centering
\begin{tabular}{|c|c|c|c|l|}
\hline
\textbf{Exchangeability} & \textbf{Misspecification} & \textbf{Notes} & \multicolumn{1}{l|}{\textbf{Scenario}} & \multicolumn{1}{c|}{\textbf{Data Generating Models}} \\ \hline
\multirow{3}{*}{\begin{tabular}[c]{@{}c@{}}No Exchangeability\\ Violation\end{tabular}} & \begin{tabular}[c]{@{}c@{}}No Model\\ Misspecification\end{tabular} & \begin{tabular}[c]{@{}c@{}}Baseline\\ Example\end{tabular}  & A & \begin{tabular}[c]{@{}l@{}}$g[\rho(\mathbf{X}_i)] = 0.5 - 0.5X_{i1} + 0.5X_{i2} - 0.5X_{i3} + 0.5X_{i4}$\\ $g[\pi_{S_i}(\mathbf{X}_i)] = 0.5X_{i1} - 0.5X_{i2} + 0.5X_{i3} - 0.5X_{i4}$\\ $\mu_0(S_i, \mathbf{X}_i) = 2 - 3X_{i1} - X_{i2} + X_{i3} + 3X_{i4}$\\ $\mu_1(S_i, \mathbf{X}_i) = \mu_0(S_i, \mathbf{X}_i) - 2 - X_{i1} + 3X_{i2} - 3X_{i3} + X_{i4}$\end{tabular} \\ \cline{2-5} 
 & \multirow{2}{*}{\begin{tabular}[c]{@{}c@{}}Outcome Model\\ Misspecification\end{tabular}} & \begin{tabular}[c]{@{}c@{}}Sample\\ Overlap\\ Violation\end{tabular} & B & \begin{tabular}[c]{@{}l@{}}$g[\rho(\mathbf{X}_i)] = 2 - 2X_{i1} + 2X_{i2} - 2X_{i3} + 2X_{i4}$\\ $g[\pi_{S_i}(\mathbf{X}_i)] = 0.5X_{i1} - 0.5X_{i2} + 0.5X_{i3} - 0.5X_{i4}$\\ $\mu_0(S_i, \mathbf{X}_i) = 2 - 3U_{i1} - U_{i2} + U_{i3} + 3U_{i4}$\\ $\mu_1(S_i, \mathbf{X}_i) = \mu_0(S_i, \mathbf{X}_i) - 2 - U_{i1} + 3U_{i2} - 3U_{i3} + U_{i4}$\end{tabular} \\ \cline{3-5} 
 &  & \begin{tabular}[c]{@{}c@{}}Treatment\\ Overlap\\ Violation\end{tabular} & C & \begin{tabular}[c]{@{}l@{}}$g[\rho(\mathbf{X}_i)] = 0.5 - 0.5X_{i1} + 0.5X_{i2} - 0.5X_{i3} + 0.5X_{i4}$\\ $g[\pi_{S_i}(\mathbf{X}_i)] = 2X_{i1} - 2X_{i2} + 2X_{i3} - 2X_{i4}$\\ $\mu_0(S_i, \mathbf{X}_i) = 2 - 3U_{i1} - U_{i2} + U_{i3} + 3U_{i4}$\\ $\mu_1(S_i, \mathbf{X}_i) = \mu_0(S_i, \mathbf{X}_i) - 2 - U_{i1} + 3U_{i2} - 3U_{i3} + U_{i4}$\end{tabular} \\ \hline
\begin{tabular}[c]{@{}c@{}}Potential Outcome\\ Exchangeability\\ Violation\end{tabular} & \begin{tabular}[c]{@{}c@{}}Sampling and \\ Propensity Score\\ Misspecification\end{tabular} & \begin{tabular}[c]{@{}c@{}}Prognostic\\ Sampling Variable\end{tabular} & D & \begin{tabular}[c]{@{}l@{}}$g[\rho(\mathbf{X}_i)] = 0.5 - 0.5U_{i1} + 0.5U_{i2} - 0.5U_{i3} + 0.5U_{i4}$\\ $g[\pi_{S_i}(\mathbf{X}_i)] = 0.5U_{i1} - 0.5U_{i2} + 0.5U_{i3} - 0.5U_{i4}$\\ $\mu_0(S_i, \mathbf{X}_i) = S_i\left(2 - 3X_{i1} - X_{i2} + X_{i3} + 3X_{i4}\right)$\\ $\qquad\qquad + (1 - S_i)\left(2X_{i1} - 2X_{i2} - 2X_{i3} + 2X_{i4}\right)$\\ $\mu_1(S_i, \mathbf{X}_i) = \mu_0(S_i, \mathbf{X}_i) - 2 - X_{i1} + 3X_{i2} - 3X_{i3} + X_{i4}$\end{tabular} \\ \hline
\begin{tabular}[c]{@{}c@{}}Propensity Score\\ Exchangeability\\ Violation\end{tabular} & \begin{tabular}[c]{@{}c@{}}Outcome Model\\ Misspecification\end{tabular} & \begin{tabular}[c]{@{}c@{}}Instrumental\\ Sampling Variable\end{tabular} & E & \begin{tabular}[c]{@{}l@{}}$g[\rho(\mathbf{X}_i)] = 0.5 - 0.5X_{i1} + 0.5X_{i2} - 0.5X_{i3} + 0.5X_{i4}$\\ $g[\pi_{S_i}(\mathbf{X}_i)] = S_i\left(0.5X_{i1} - 0.5X_{i2} + 2X_{i3} - 2X_{i4}\right) - 0.5(1 - S_i)$\\ $\mu_0(S_i, \mathbf{X}_i) = 2 - 3U_{i1} - U_{i2} + U_{i3} + 3U_{i4}$\\ $\mu_1(S_i, \mathbf{X}_i) = \mu_0(S_i, \mathbf{X}_i) - 2 - U_{i1} + 3U_{i2} - 3U_{i3} + U_{i4}$\end{tabular} \\ \hline
\multirow{3}{*}{\begin{tabular}[c]{@{}c@{}}Propensity Score and \\ Potential Outcome\\ Exchangeability\\ Violation\end{tabular}} & \begin{tabular}[c]{@{}c@{}}No Model \\ Misspecification\end{tabular} & \multirow{3}{*}{\begin{tabular}[c]{@{}c@{}}Confounding Sampling\\ Variable\end{tabular}} & F & \begin{tabular}[c]{@{}l@{}}$g[\rho(\mathbf{X}_i)] = 0.5 - 0.5X_{i1} + 0.5X_{i2} - 0.5X_{i3} + 0.5X_{i4}$\\ $g[\pi_{S_i}(\mathbf{X}_i)] = S_i\left(0.5X_{i1} - 0.5X_{i2} + 0.5X_{i3} - 0.5X_{i4}\right) - 0.5(1 - S_i)$\\ $\mu_0(S_i, \mathbf{X}_i) = S_i\left(2 - 3X_{i1} - X_{i2} + X_{i3} + 3X_{i4}\right)$\\ $\qquad\qquad + (1 - S_i)\left(2X_{i1} - 2X_{i2} - 2X_{i3} + 2X_{i4}\right)$\\ $\mu_1(S_i, \mathbf{X}_i) = \mu_0(S_i, \mathbf{X}_i) - 2 - X_{i1} + 3X_{i2} - 3X_{i3} + X_{i4}$\end{tabular} \\ \cline{2-2} \cline{4-5} 
 & \begin{tabular}[c]{@{}c@{}}Sampling and \\ Propensity Score\\ Misspecification\end{tabular} &  & G & \begin{tabular}[c]{@{}l@{}}$g[\rho(\mathbf{X}_i)] = 0.5 - 0.5U_{i1} + 0.5U_{i2} - 0.5U_{i3} + 0.5U_{i4}$\\ $g[\pi_{S_i}(\mathbf{X}_i)] = S_i\left(0.5U_{i1} - 0.5U_{i2} + 0.5U_{i3} - 0.5U_{i4}\right) - 0.5(1 - S_i)$\\ $\mu_0(S_i, \mathbf{X}_i) = S_i\left(2 - 3X_{i1} - X_{i2} + X_{i3} + 3X_{i4}\right)$\\ $\qquad\qquad + (1 - S_i)\left(2X_{i1} - 2X_{i2} - 2X_{i3} + 2X_{i4}\right)$\\ $\mu_1(S_i, \mathbf{X}_i) = \mu_0(S_i, \mathbf{X}_i) - 2 - X_{i1} + 3X_{i2} - 3X_{i3} + X_{i4}$\end{tabular} \\ \cline{2-2} \cline{4-5} 
 & \begin{tabular}[c]{@{}c@{}}Outcome Model\\ Misspecification\end{tabular} &  & H & \begin{tabular}[c]{@{}l@{}}$g[\rho(\mathbf{X}_i)] = 0.5 - 0.5X_{i1} + 0.5X_{i2} - 0.5X_{i3} + 0.5X_{i4}$\\ $g[\pi_{S_i}(\mathbf{X}_i)] = S_i\left(0.5X_{i1} - 0.5X_{i2} + 0.5X_{i3} - 0.5X_{i4}\right) - 0.5(1 - S_i)$\\ $\mu_0(S_i, \mathbf{X}_i) = S_i\left(2 - 3U_{i1} - U_{i2} + U_{i3} + 3U_{i4}\right)$\\ $\qquad\qquad + (1 - S_i)\left(2U_{i1} - 2U_{i2} - 2U_{i3} + 2U_{i4}\right)$\\ $\mu_1(S_i, \mathbf{X}_i) = \mu_0(S_i, \mathbf{X}_i) - 2 - U_{i1} + 3U_{i2} - 3U_{i3} + U_{i4}$\end{tabular} \\ \hline
\end{tabular}
\caption{Simulation scenarios and the corresponding generative models for $\rho(\mathbf{X}_i)$, $\pi_s(\mathbf{X}_i)$, $\mu_0(\mathbf{X}_i)$, and $\mu_1(\mathbf{X}_i)$.}\label{generative:04}
\end{table}

In the transportability examples, we will discard both $Y_i$ and $Z_i$ for all $\{i:S_i = 0\}$, and use only $Y_i$ and $Z_i$ from units $\{i:S_i = 1\}$. The data-fusion methods will make use of $Y_i$ and $Z_i$ for all $i = 1,2,\ldots,n$. Regardless of whether the method estimates $\tau_0$ under the transportability case or the data-fusion case, each estimator is provided a design matrix containing an intercept term and the four original covariate measurements - $X_{i1}$, $X_{i2}$, $X_{i3}$, and $X_{i4}$ for all $i = 1,2,\ldots,n$. The target population average treatment effect is estimated across each iteration of a given scenario using the respective estimators described in Section \ref{transport:04}: $\hat{\tau}_{\text{TMLE}}$, $\hat{\tau}_{\text{AUG}}$ and $\hat{\tau}_{\text{CAL}}$. We also find estimates of $\tau_0$ using the data-fusion estimators $\tilde{\tau}_{\text{AUG}}$ and $\tilde{\tau}_{\text{CAL}}$ from Section \ref{fusion:04}.

\subsection{Simulation Results}\label{sim-results:04}

\begin{figure}[H]
	\centering
	\includegraphics[scale = 0.12]{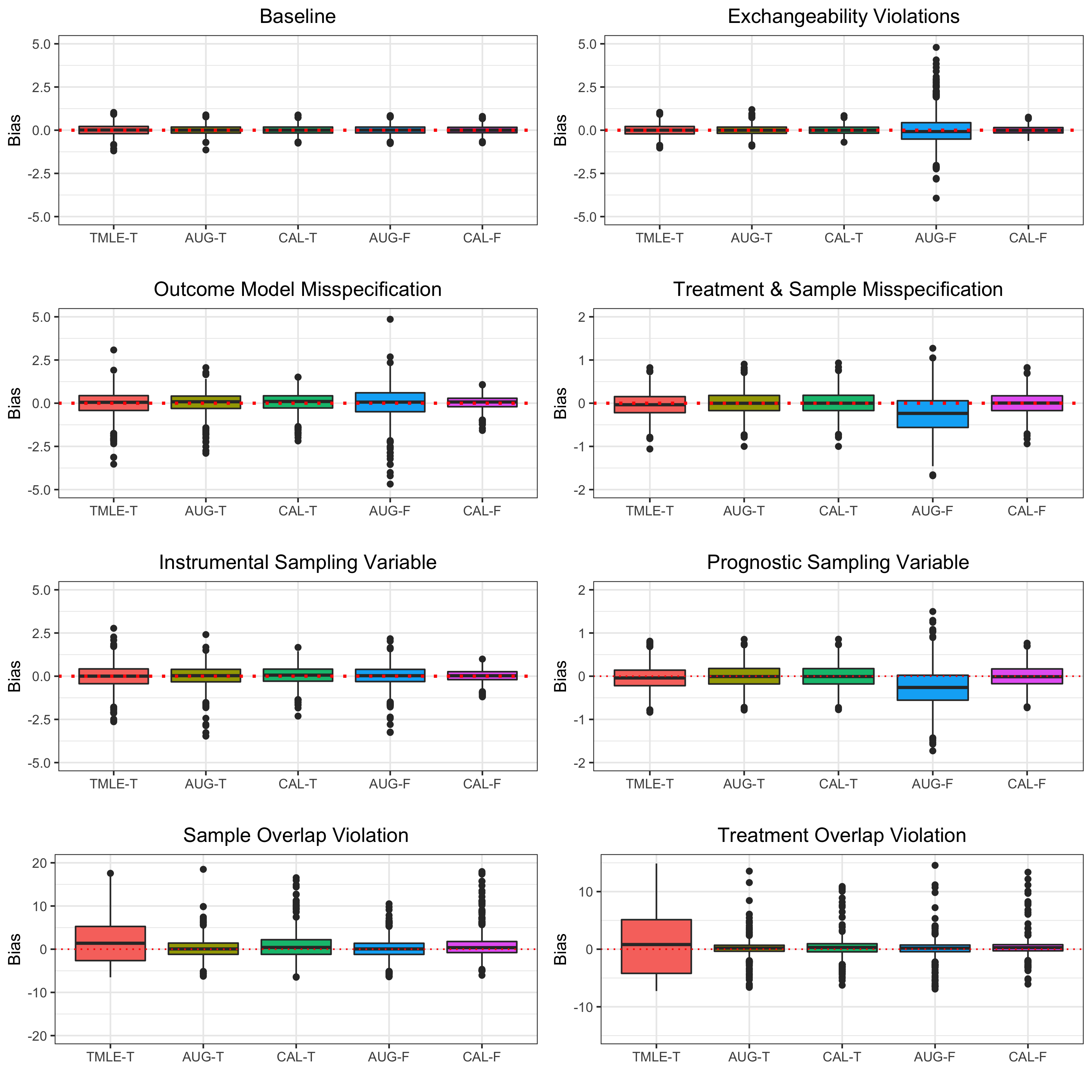}
	\caption{A subset of the constant conditional ATE estimates using four different methods for estimating balancing weights. Each boxplot is composed of $1{,}000$ estimates from the replicates that generate the values in Table \ref{target-table:04}. The suffixes -T denotes the transportability case, while -F denotes the data-fusion case.}\label{ATE-plot:04}
\end{figure}

We report the average bias and root mean square error (RMSE) for each of the scenarios described in Table \ref{generative:04} across $1{,}000$ iterations using the estimators described in Sections \ref{transport:04} and \ref{fusion:04}. The results of the experiment are summarized in Table \ref{target-table:04} for $n = 500$ and $n = 2{,}000$. Figure \ref{ATE-plot:04} provides a graphical representation of the results for $n = 1{,}000$. We also report the coverage probabilities in Table \ref{coverage-table:04}. The augmented and TMLE approaches use a plug-in estimator of the efficient influence function to find the 95\% confidence intervals \cite{rudolph_robust_2017}. For the full calibration approaches, the 95\% confidence interval is estimated using a robust variance estimator \citep{stefanski_calculus_2002}.

\begin{table}[H]
\footnotesize
\centering
\begin{tabular}{cccccccccc}
\hline
\multirow{2}{*}{$n$} & \multirow{2}{*}{Scenario} & \multirow{2}{*}{$\tau_0$} &  & \multicolumn{3}{c}{Transportability} &  & \multicolumn{2}{c}{Data-Fusion} \\ \cline{5-7} \cline{9-10} 
 &  &  &  & TMLE & AUG & CAL &  & AUG & CAL \\ \hline
500 & A & -4.00 &  & -0.01 (0.45) & 0.00 (0.38) & -0.01 (0.36) &  & 0.00 (0.38) & 0.00 (0.32) \\
500 & B & -3.51 &  & 2.73 (6.10) & 0.38 (2.69) & 1.95 (4.69) &  & 0.37 (2.81) & 2.70 (5.43) \\
500 & C & -2.71 &  & 1.33 (5.50) & 0.27 (1.61) & 0.57 (2.54) &  & 0.19 (1.73) & 0.94 (3.50) \\
500 & D & -2.73 &  & -0.05 (0.37) & -0.01 (0.35) & -0.01 (0.35) &  & -0.27 (0.72) & -0.01 (0.33) \\
500 & E & -2.71 &  & -0.03 (0.95) & 0.00 (0.83) & 0.04 (0.72) &  & 0.00 (0.86) & 0.03 (0.50) \\
500 & F & -4.00 &  & 0.02 (0.46) & 0.01 (0.37) & 0.02 (0.36) &  & 0.03 (1.35) & 0.01 (0.32) \\
500 & G & -2.73 &  & -0.03 (0.38) & 0.01 (0.36) & 0.01 (0.36) &  & -0.25 (0.70) & 0.00 (0.34) \\
500 & H & -2.71 &  & -0.01 (1.02) & 0.02 (0.83) & 0.06 (0.78) &  & 0.04 (1.11) & 0.03 (0.54) \\
 &  &  &  &  &  &  &  &  &  \\
2000 & A & -4.00 &  & -0.01 (0.22) & 0.00 (0.19) & 0.00 (0.18) &  & 0.00 (0.19) & 0.00 (0.16) \\
2000 & B & -3.51 &  & 0.85 (4.78) & 0.09 (1.63) & 0.25 (2.27) &  & 0.06 (1.67) & 0.31 (2.25) \\
2000 & C & -2.71 &  & 0.31 (4.59) & 0.11 (1.09) & 0.13 (0.92) &  & 0.13 (1.32) & 0.15 (0.67) \\
2000 & D & -2.73 &  & -0.04 (0.19) & 0.00 (0.17) & 0.00 (0.17) &  & -0.24 (0.39) & 0.00 (0.17) \\
2000 & E & -2.71 &  & -0.04 (0.48) & -0.01 (0.46) & 0.03 (0.38) &  & -0.01 (0.45) & 0.01 (0.26) \\
2000 & F & -4.00 &  & -0.01 (0.21) & 0.00 (0.19) & 0.00 (0.18) &  & -0.01 (0.60) & 0.00 (0.16) \\
2000 & G & -2.73 &  & -0.04 (0.20) & 0.00 (0.19) & 0.00 (0.19) &  & -0.26 (0.40) & 0.00 (0.18) \\
2000 & H & -2.71 &  & -0.01 (0.47) & 0.02 (0.43) & 0.05 (0.38) &  & 0.01 (0.65) & 0.03 (0.26) \\ \hline
\end{tabular}
\caption{Average Bias (RMSE) of the various Transportability and Data-Fusion estimators under several different simulation scenarios of the treatment assignment, sampling and outcome processes described in Table \ref{generative:04}.}\label{target-table:04}
\end{table}

In the case of transportability, it is clear that the three methods from Section \ref{transport:04} perform equitably in scenarios where model misspecification occurs, and when there is some superseding exchangeability violation (Assumptions \ref{out-exchange:04} and \ref{ps-exchange:04}). The former scenarios demonstrate that each method is doubly-robust whereas the the latter scenarios make it abundantly clear that transportability is not affected by the potential outcome exchangeability assumption nor the propensity score exchangeability assumption. The full calibration approach achieved a lower RMSE whenever the outcome model was misspecified (Scenarios E and H). When the potential outcome models are correctly specified, the trade-offs between the different methods for transporting observational results are negligible. Regarding the scenarios with practical violations to overlap, it is apparent that the full calibration approach was less vulnerable in scenarios with treatment overlap violation (Scenario C) relative to the scenarios in which sample overlap is violated (Scenario B). The augmented approach achieved the least amount of bias in scenarios where there is limited sample overlap in terms of the average bias and the RMSE. When $n = 500$, the augmented approach performed better than the full calibration approach in scenarios where treatment group overlap was violated (Scenario C), but when $n = 2{,}000$ the degree of bias between the two methods is nearly identical. In this latter case, the full calibration approach had a smaller RMSE. Even though the full calibration approach performed worse than the augmented approach in scenarios involving a sample overlap violation, sample heterogeneity is still resolved using calibration thus marking the importance of this type of approach. This claim is further evidenced by the relatively poor results generated by the TMLE estimator in these settings. Furthermore, in Table \ref{coverage-table:04}, it appears that the full calibration approach retains accurate coverage probabilities of $\tau_0$, whereas finding the 95\% confidence interval with a plug-in estimator of the influence function sometimes overestimates the variance of the target population average treatment effect estimates.

\begin{table}[H]
\footnotesize
\centering
\begin{tabular}{cccccccccc}
\hline
\multirow{2}{*}{$n$} & \multirow{2}{*}{Scenario} & \multirow{2}{*}{$\tau_0$} &  & \multicolumn{3}{c}{Transportability} &  & \multicolumn{2}{c}{Data-Fusion} \\ \cline{5-7} \cline{9-10} 
 &  &  &  & TMLE & AUG & CAL &  & AUG & CAL \\ \hline
500 & A & -4.00 &  & 0.946 & 0.946 & 0.929 &  & 0.954 & 0.940 \\
500 & B & -3.51 &  & 0.371 & 0.843 & 0.625 &  & 0.913 & 0.625 \\
500 & C & -2.71 &  & 0.134 & 0.898 & 0.690 &  & 0.903 & 0.694 \\
500 & D & -2.73 &  & 0.968 & 0.956 & 0.957 &  & 0.995 & 0.956 \\
500 & E & -2.71 &  & 0.899 & 0.952 & 0.927 &  & 0.976 & 0.951 \\
500 & F & -4.00 &  & 0.942 & 0.949 & 0.939 &  & 0.991 & 0.947 \\
500 & G & -2.73 &  & 0.970 & 0.958 & 0.960 &  & 0.993 & 0.957 \\
500 & H & -2.71 &  & 0.867 & 0.941 & 0.899 &  & 0.991 & 0.943 \\
 &  &  &  &  &  &  &  &  &  \\
2000 & A & -4.00 &  & 0.960 & 0.954 & 0.951 &  & 0.953 & 0.951 \\
2000 & B & -3.51 &  & 0.325 & 0.962 & 0.686 &  & 0.967 & 0.706 \\
2000 & C & -2.71 &  & 0.113 & 0.917 & 0.736 &  & 0.927 & 0.774 \\
2000 & D & -2.73 &  & 0.964 & 0.957 & 0.957 &  & 0.988 & 0.956 \\
2000 & E & -2.71 &  & 0.910 & 0.953 & 0.930 &  & 0.957 & 0.948 \\
2000 & F & -4.00 &  & 0.962 & 0.946 & 0.941 &  & 0.987 & 0.949 \\
2000 & G & -2.73 &  & 0.948 & 0.937 & 0.934 &  & 0.977 & 0.932 \\
2000 & H & -2.71 &  & 0.903 & 0.950 & 0.934 &  & 0.979 & 0.957 \\ \hline
\end{tabular}
\caption{Coverage probabilities of $\tau_0$ for the methods in Section \ref{transport:04} and \ref{fusion:04}.}\label{coverage-table:04}
\end{table}

When we combine datasets for data-fusion, we might expect to get more precise estimates of $\tau_0$ than with an analogous transportability estimator. However, this is not the case for the data-fusion variant to the augmented approach. In many scenarios, the augmented estimator for data-fusion had a higher RMSE than the augmented estimator for transporting observational results. Recall that the proposed augmented solution has the same estimator in the data-fusion case as in the transportability case. The only difference between the two is the choice of initial outcome model. The only time that the RMSE is smaller was when some overlap violation occurs. The only occurrence where the full calibration approach to data-fusion was more accurate than the augmented approach concerning poor overlap conditions was when the treatment overlap assumption was violated (Scenario C) and $n = 2{,}000$. The same result occurs for the transportability case. Moreover, $\tilde{\tau}_{\text{AUG}}$ is biased whenever potential outcome exchangeability is violated and the propensity and sampling scores are incorrectly specified (Scenarios D and G), whereas the transportability variant of the augmented estimator is unbiased in these same scenarios. Once again, the coverage probabilities are accurate for the full calibration approach to data-fusion, yet the variance is over estimated for the augmented approach as seen in Table \ref{coverage-table:04}.

\section{Evaluating Metformin Versus Sulfonylureas as Monotherapy for VA Diabetes Patients}\label{illustrate:04}

The primary aim of this applied analysis is to estimate the risk difference in mortality between sulfonylurea and metformin monotherapy for the 2010-2014 cohort of diabetic VA patients. We first use improved covariate balance propensity score (CBPS) weights \citep{fan_improving_2018, josey2021framework} to estimate the risk difference using only the 2010-2014 cohort, which balance the covariates found in Table \ref{table-one:04}. This estimate will serve as a benchmark for the transport and data-fusion settings. We also estimate the risk difference for the 2004-2009 cohort without the 2010-2014 data, again for comparative purposes. We use (\ref{dual-cal:04})-(\ref{hajek:04}) to find estimates of total mortality in the 2004-2009 sample transported to the 2010-2014 cohort. Since sulfonylurea use in 2004-2009 represents 27.0\% of monotherapy recipients but only 11.8\% of patients in 2010-2014 implies the need for data-fusion methods which account for propensity score exchangeability violations, like that of the estimator in (\ref{dual-fusion:04})-(\ref{hajek-fusion:04}).

Both cohorts excluded patients with pre-existing forms of cancer. We also omitted patients that received a second-line medication (either insulin, a thiazolidinedione, a sulfonylurea for patients receiving metformin, or metformin for patients receiving a sulfonylurea) within 30 days of their first filled prescription of either metformin or a sulfonylurea. Time to mortality, which is used to create the indicators of the three mortality outcomes, is computed as the number of days to death from the date when the first prescription is filled. Our analysis assumes intention to treat with the first prescribed therapy. We do not censor patients at the time when a second-line medication is prescribed, as was done in \cite{wheeler_mortality_2013}. The remaining baseline demographic and clinical characteristics about the two cohorts are summarized in Table \ref{table-one:04}. 

Analyses based on covariate balance methods typically require checks to evaluate whether balance is actually achieved, both between the two samples and between the two treatment groups under investigation. A common diagnostic statistic for measuring balance is the standardized absolute mean difference evaluated over the various covariate measurements after weighting \citep{austin2009balance}. We report this statistic for the unweighted differences in Table \ref{table-one:04}. With calibration, this step is redundant because the calibration weights should attain a standardized mean difference identical to zero (disregarding computational tolerance) in every feasible occasion (i.e. when a solution to the desired primal problem exists). However, it can still be helpful from a pedagogical standpoint to visualize this result. In the online supplement, we analyze a similar dataset comparing insulin sensitization therapy with insulin provisioning therapy for Type 2 diabetes patients -- transporting and combining the results of the BARI 2D study \citep{the_bari_2d_study_group_randomized_2009} with the 2010-2014 VA cohort. There we show a plot that illustrates the exact balancing nature of the calibration method, both between treatments and samples evaluated with the weighted standardized absolute mean differences. This supplemental analysis also provides an example for how we can combine experimental and observational datasets for data-fusion.

\begin{figure}[H]
	\centering
	\includegraphics[scale = 0.5]{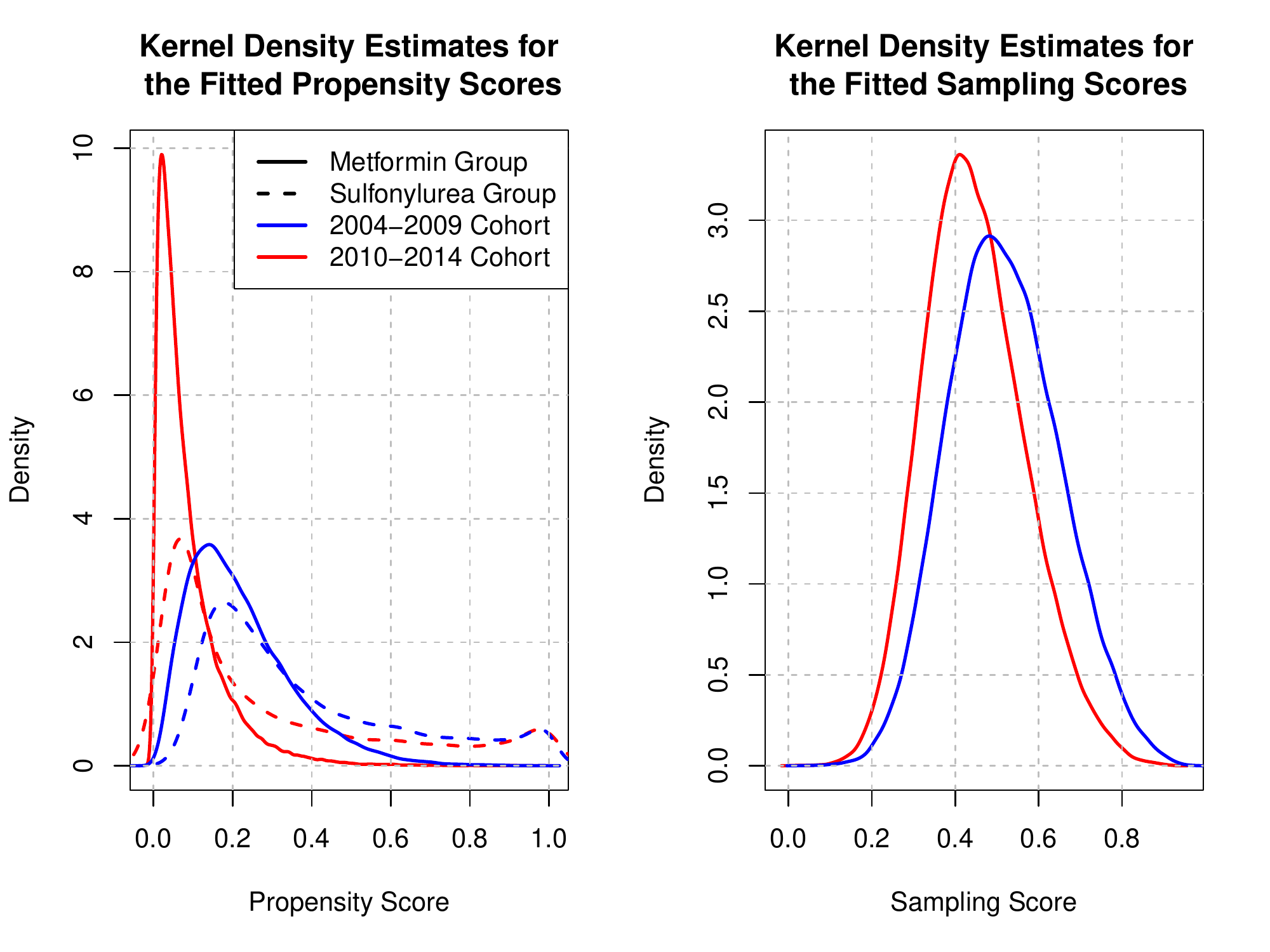}
	\caption{Kernel density estimates of the CBPS \citep{fan_improving_2018} fitted propensity and sampling scores. Recall that the propensity score measures the probability of being assigned to a sulfonylurea and the sampling score measures the probability of belonging to the 2004-2009 cohort versus the 2010-2014 cohort.}\label{overlap:04}
\end{figure}

Another diagnostic is to compare the estimated propensity and sampling scores to ensure that there is sufficient overlap between the two treatment groups and between the two samples. We use a nonparametric kernel density estimator of the probability density function for the fitted propensity and sampling scores to visualize the degree that the treatment and sample stratified distributions overlap. Both the propensity score and the sampling score were fit using covariate balance propensity scores \citep{fan_improving_2018}. This diagnostic reveals the validity of the overlap assumptions (Assumptions \ref{positivity-sample:04} and \ref{positivity-treat:04}). In Figure \ref{overlap:04} we can see that there is indeed broad overlap between the sulfonylurea and metformin group, both in the 2004-2009 cohort and in the 2010-2014 cohort, and a considerable amount of overlap exists between samples based on the predicted sampling scores. Note that it is typically not possible to visualize the degree of overlap between treatments in the target sample under the transportability scenario, unless the treatment assignment is observed. Since our example works for both the transportability and data-fusion scenarios, we are able to estimate the propensity score in the target sample displayed in Figure \ref{overlap:04}.

Table \ref{risk-diff:04} contains the various risk-difference estimates that we computed on the illustrative dataset. We will primarily focus on the estimates of five-year mortality since these figures have the greatest magnitude. Furthermore, the trends that we will report about five-year mortality appear to be the same for one- and two-year mortality. Observe that the crude risk difference in five-year mortality is similar within both the 2004-2009 and in the 2010-2014 cohorts. This implies that there is either limited or no changes to the risk difference attributable to temporal trends in treatment effectiveness. Note that a temporal effect modifier is one factor that we would not be able to accommodate without violating sample positivity as it is collinear with the sample indicator. The adjusted marginal estimates of the risk difference using CBPS reveal the importance of accounting for differences observed between the study and target cohorts. The risk difference in the 2004-2009 cohort is 12.2\%, 95\% CI = (11.6\%, 12.7\%), and 4.1\%, 95\% CI = (3.4\%, 4.9\%), in the 2010-2014 cohort. Given the limited change of the crude estimates between cohorts, these discrepancies are likely due to differences in the distribution of effect modifiers between the two temporally-distinct cohorts. When we transport the estimates of the 2004-2009 cohort onto the 2010-2014 cohort, the risk difference is found to be 4.0\%, 95\% CI = (3.4\%, 4.6\%). That is, the transported effect estimate is more similar to the CBPS estimate of 2010-2014 than of the CBPS estimate evaluated on the 2004-2009 cohort. We also see that the transported estimate from the 2004-2009 cohort onto the 2010-2014 cohort is more efficient than the calibrated estimate computed with only the data from the 2010-2014 cohort.

\begin{table}[H]
\centering
\scriptsize
\begin{tabular}{lccccc}
\hline
 & \begin{tabular}[c]{@{}c@{}}Metformin\\ (2004-2009)\end{tabular} & \begin{tabular}[c]{@{}c@{}}Metformin\\ (2010-2014)\end{tabular} & \begin{tabular}[c]{@{}c@{}}Sulfonylurea\\ (2004-2009)\end{tabular} & \begin{tabular}[c]{@{}c@{}}Sulfonylurea\\ (2010-2014)\end{tabular} & \begin{tabular}[c]{@{}c@{}}Standardized Absolute\\ Mean Difference\end{tabular} \\ \hline
Patient Count & 84003 & 100612 & 29447 & 11736 & - \\
Baseline Age & 61.90 (11.64) & 60.45 (11.50) & 67.32 (12.47) & 66.95 (12.72) & 0.354 \\
Male & 80964 (96.4) & 95906 (95.3) & 28912 (98.2) & 11472 (97.8) & 0.095 \\
Race/Ethnicity &  &  &  &  & 0.185 \\
\multicolumn{1}{r}{Non-hispanic White} & 11927 (14.2) & 20132 (20.0) & 4038 (13.7) & 2253 (19.2) &  \\
\multicolumn{1}{r}{Non-hispanic Black} & 4730 (5.6) & 6810 (6.8) & 1517 (5.2) & 614 (5.2) &  \\
\multicolumn{1}{r}{Hispanic} & 10897 (13.0) & 8583 (8.5) & 5073 (17.2) & 1123 (9.6) &  \\
\multicolumn{1}{r}{Other} & 56449 (67.2) & 65087 (64.7) & 18819 (63.9) & 7746 (66.0) &  \\
Smoking Status &  &  &  &  & 0.127 \\
\multicolumn{1}{r}{Current} & 21475 (25.6) & 30603 (30.4) & 6365 (21.6) & 2925 (24.9) &  \\
\multicolumn{1}{r}{Former} & 42751 (50.9) & 44354 (44.1) & 16396 (55.7) & 5935 (50.6) &  \\
\multicolumn{1}{r}{Never} & 19777 (23.5) & 25655 (25.5) & 6686 (22.7) & 2876 (24.5) &  \\
BMI & 32.96 (6.43) & 33.66 (6.52) & 31.20 (6.00) & 32.08 (6.22) & 0.220 \\
SBP & 133.03 (16.61) & 132.35 (15.93) & 133.83 (18.37) & 131.80 (17.31) & 0.066 \\
DBP & 76.83 (10.82) & 78.69 (10.76) & 74.90 (11.59) & 75.81 (11.45) & 0.185 \\
HDL & 39.19 (10.76) & 40.14 (10.97) & 39.23 (11.51) & 39.60 (11.50) & 0.048 \\
LDL & 103.43 (34.93) & 102.68 (35.30) & 101.32 (35.46) & 96.99 (35.06) & 0.098 \\
Total Cholesterol & 179.60 (44.15) & 178.57 (44.51) & 177.91 (46.41) & 173.49 (46.68) & 0.069 \\
Triglycerides & 205.36 (192.15) & 203.67 (198.23) & 206.01 (198.88) & 208.68 (216.49) & 0.013 \\
Fasting Plasma Glucose & 151.54 (62.26) & 149.71 (63.64) & 165.87 (80.76) & 163.67 (78.23) & 0.141 \\
HbA1c & 7.17 (1.43) & 7.28 (1.41) & 7.42 (1.63) & 7.54 (1.54) & 0.139 \\
Estimated GFR & 78.41 (18.59) & 83.88 (20.08) & 66.83 (22.97) & 66.15 (24.76) & 0.500 \\
Creatinine & 1.03 (0.20) & 0.98 (0.20) & 1.25 (0.48) & 1.29 (0.56) & 0.493 \\
History of CAD & 31211 (37.2) & 25771 (25.6) & 14071 (47.8) & 4535 (38.6) & 0.240 \\
History of CHF & 8768 (10.4) & 6086 (6.0) & 5916 (20.1) & 1752 (14.9) & 0.237 \\
History of Stroke & 10572 (12.6) & 8385 (8.3) & 5230 (17.8) & 1645 (14.0) & 0.149 \\
History of Kidney Disease & 571 (0.7) & 237 (0.2) & 1214 (4.1) & 316 (2.7) & 0.167 \\
History of Liver Disease & 424 (0.5) & 349 (0.3) & 278 (0.9) & 86 (0.7) & 0.043 \\ \hline
\end{tabular}
\caption{Summary statistics for covariates measured on newly diagnosed diabetic patients receiving care in the VA healthcare system stratified by years (2004-2009 and 2010-2014) and monotherapy type (Metformin or Sulfonylurea). The standardized absolute mean differences is an average measurement between the four temporal and treatment defined groups.}\label{table-one:04}
\end{table}

We also examined the scenario which combines the 2004-2009 sample with the 2010-2014 sample to estimate the risk difference among patients in the 2010-2014 cohort using (\ref{dual-fusion:04})-(\ref{hajek-fusion:04}), which accommodates the data-fusion setting. When we combine the two cohorts, we get an unbiased estimate of $\tau_0$ with much greater efficiency than using only data from 2010-2014 alone. The efficiency gain is due to the increased effective sample size after weighting, which increased modestly from 30{,}883 in the transportability case to 37{,}682 in the data-fusion case. Using the data-fusion calibration estimator, we found the estimated risk difference for 2010-2014 to be 4.2\%, 95\% CI = (3.7\%, 4.7\%). Regardless of the estimator, our results suggest that metformin monotherapy remains associated with lower mortality than sulfonylurea treatment for veterans with newly-diagnosed diabetes. However, the size of the effect has declined over the two time periods, potentially implying improved patient selection for sulfonylurea treatment.

\begin{table}[H]
\footnotesize
\centering
\begin{tabular}{cccc}
\hline
Method and Sample & One Year after Rx & Two Years after Rx & Five Years after Rx \\ \hline
Unadjusted 2004-2009 & 2.6\% (2.3\%, 2.8\%) & 5.3\% (4.9\%,5.7\%) & 12.6\% (11.9\%, 13.2\%) \\
Unadjusted 2010-2014 & 2.4\% (2.0\%, 2.7\%) & 5.1\% (4.6\%, 5.7\%) & 12.5\% (11.7\%, 13.4\%) \\
CBPS 2004-2009 & 2.3\% (2.1\%, 2.6\%) & 5.0\% (4.7\%, 5.4\%) & 12.2\% (11.6\%, 12.7\%) \\
CBPS 2010-2014 & 1.0\% (0.7\%, 1.4\%) & 2.0\% (1.5\%, 2.5\%) & 4.1\% (3.4\%, 4.9\%) \\
Transported 2004-2009 & 0.8\% (0.5\%, 1.1\%) & 1.9\% (1.4\%, 2.3\%) & 4.0\% (3.4\%, 4.6\%) \\
Data-Fusion & 0.9\% (0.7\%, 1.1\%) & 2.1\% (1.7\%, 2.4\%) & 4.2\% (3.7\%, 4.7\%) \\ \hline
\end{tabular}
	\caption{Risk differences in total mortality between sulonylurea and metformin monotherapy in a VA cohort starting from the date of first prescription (Rx) using a variety of causal effect estimators.}\label{risk-diff:04}
\end{table}

\section{Discussion}\label{discussion:04}

By estimating a vector of weights that simultaneously reduces sample and treatment group heterogeneity, we show how constrained convex optimization techniques can be applied to synergize solutions for causal inference using covariate balancing weights \citep{josey2021framework} with solutions for transporting experimental effect estimates using sampling weights \citep{signorovitch_comparative_2010, westreich_transportability_2017}. The resulting estimator eliminates both confounding and sampling bias. This allows us to transport estimates found with observational data across populations. We also examined two alternative approaches for transporting experimental results which we adapted to accommodate data from observational studies. The TMLE \citep{rudolph_robust_2017} and augmented estimators \citep{lee2021improving} are less constrained by parametric assumptions, but their operational qualities are less interpretable if the desired objective is to balance covariate distributions between samples and treatment groups. We then demonstrate how the proposed full calibration approach can be extended to solve the problem of data-fusion in an observational context.

In the simulation study conducted in Section \ref{simulation:04}, we found that using some form of calibration, either with the augmented approach or the full calibration approach, yielded the most efficient estimates. In cases where the potential outcome models are correctly specified, the difference in performance between the three transportability estimators was negligible. For data-fusion, we saw in the simulation that the full calibration approach outperformed the alternative augmented approach. It is arguable that the increased efficiency might be the result of the implied parametric nature of the full calibration approaches. However, in this simulation, the nuisance parameters used by the TMLE and augmented estimators were fit using parametric models in an effort to better contrast the results. In addition to the simulation study, we show in an illustrative example evaluating mortality rates in US veterans with diabetes how different populations produce different effect estimates for the same outcome. We then demonstrate how eliminating the sampling bias induced by differences in the distribution of the effect modifiers between cohorts along with eliminating confounder heterogeneity between treatment groups produces consistent estimates of the treatment effect on the target population. 

One of the major shortcomings of the full calibration method is the set of linearity conditions nested within Assumptions \ref{linear:04}-\ref{odds-treat:04}. These assumptions are necessary to guarantee the double-robustness property for this estimator. The TMLE approach does not require any assumption about the functional form of the nuisance parameters - the only requirement is that the nuisance parameter models be statistically consistent. The augmented approach as presented in Section \ref{augment:04} only requires Assumption \ref{odds-sample:04}, along with a statistically consistent propensity score, or consistent potential outcome models to guarantee double-robustness. In data-fusion, we show that Assumption \ref{out-exchange:04} is required for the augmented approach in instances where the sampling and propensity scores are not correctly specified. We also note that the more stringent linearity assumptions culminate in a trade-off between more flexible modeling strategies and an increase to efficiency in finite sample settings, as demonstrated in the simulation study.

For much of this paper we have left the specification of the balancing weights $c_j(\mathbf{X}_i)$ to be ill-defined other than to set $c_1(\mathbf{X}_i) = 1$ for all $i = 1,2,\ldots,n$. This latter property is necessary to stabilize the weights. Indeed, for many problems the choice of $c_j(\mathbf{X}_i)$ can simply be the original covariate measurements without any transformation, as we demonstrated in the simulations and illustrative examples. A more flexible yet complex option is to use sieve-type regression methods \citep{geman_nonparametric_1982} that replace the balance functions in the dual problems of (\ref{dual-eb:04}), (\ref{dual-cal:04}) and (\ref{dual-fusion:04}) with an expanding basis of functions and interactions among the covariate measurements. A sieve regression alternative would relax the linearity conditions in Assumptions \ref{linear:04}-\ref{odds-treat:04}. This nonparametric approach was explored in \cite{chan2016globally} for estimating balancing weights to mitigate confounding bias and in Lee \cite{lee2021improving} for generalizing treatment effects.

In the case where we need to balance an exceptionally large number of balance functions, there is little to no assurance that a feasible solution to the primal problems in (\ref{primal-eb:04}), (\ref{primal-cal:04}) and (\ref{primal-fusion:04}) exists. To counteract this issue, we can introduce a penalization term on the dual variables in (\ref{dual-eb:04}), (\ref{dual-cal:04}) and (\ref{dual-fusion:04}). Doing so induces inequality constraints onto (\ref{primal-eb:04}), (\ref{primal-cal:04}) and (\ref{primal-fusion:04}), respectively, as opposed to the equality constraints that we currently solve. Introducing a penalization component, like in ridge regression or LASSO, results in a bias-variance trade-off in a finite sample setting. In finite sample settings without exact balance, estimates of $\tau_0$ might be biased when using weights derived with penalization. However, a slightly biased estimate is certainly better than no estimate at all. Using such an approach was not necessary for either the simulation scenarios that we considered or for the illustrative examples, as these two applications are not high-dimensional cases. Using inequality constraints can also be used to reduce the mean squared error of an estimate for $\tau_0$ if that is an objective of the analysis. More information on using inequality constraints with calibration can be found in \cite{wang2020minimal} for finding balancing weights and in \cite{lu2021you} for transporting effect estimates.

Early on, we stated that the complete individual-level covariate data were required for both samples. It would be advantageous to only require the marginal moment values of the covariate distribution (or balance functions) in the target sample for transporting observational study results, as these entries are often found in the scientific and medical literature in a so-called Table 1. This setting is discussed in more detail by \cite{josey2021transporting} for transporting randomized clinical trial results. In that article, we point out that any resulting inference in such a setting would involve the target \textit{sample} average treatment effect, $\tau'_0 \equiv n_{0}^{-1} \sum_{\{i:S_i = 0\}} Y_i(1) - Y_i(0)$, instead of the target \textit{population} estimand of $\tau_0$. The findings of \cite{josey2021transporting} can easily be translated to the full calibration approach we propose for transporting observational results in Section \ref{calibration:04} given that the two estimators of the target population average treatment effect are nearly identical. The augmented and TMLE approaches cannot be resolved in this context without additional assumptions given the need to sum the estimates of the potential outcome means over the target sample, which in general requires individual-level covariate data from the target sample. Given this finding, we believe calibration weights might be integrated or combined with the alternative estimators presented in this manuscript to further enhance these solutions when they are hindered by various modelling challenges, such as the setting where individual-participant study data are available but only marginal moment values of the target sample.

\section*{Acknowledgements} Kevin P. Josey was supported in part by the National Institute of Environmental Health Sciences, NIEHS 5T32ES007142, Fan Yang was supported in part by the National Science Foundation, NSF SES-1659935, Debashis Ghosh was supported in part by the National Science Foundation, NSF DMS-1914937, and Sridharan Raghavan was supported in part by the US Department of Veterans Affairs Award IK2-CX001907-01. 

\section*{Disclaimer} This manuscript was submitted to the Department of Biostatistics and Informatics in the Colorado School of Public Health, University of Colorado Anschutz Medical Campus, in partial fulfillment of the requirements for the degree of Doctor of Philosophy in Biostatistics for Kevin Josey.

\section*{Supplementary Material}

The R package used to fit balancing and sampling weights is in development with a working version available at https://github.com/kevjosey/cbal/. The code used to conduct the simulation study in Section \ref{simulation:04} is available at the following URL: https://github.com/kevjosey/fusion-sim/.

BARI 2D study data is publicly available through the US National Institutes of Health, National Heart, Lung and Blood Institute’s Biologic Specimen and Data Repository Information Coordinating Center (https://biolincc.nhlbi.nih.gov/studies/bari2d/). VA diabetes patient data included in this study are available on reasonable request to SR and upon obtaining required regulatory approvals according to current VA guidelines. Due to the sensitivity of the clinical data collected for this study, data requests must be from qualified researchers with approved human subjects research protocols.

\bibliographystyle{apalike}
\bibliography{fbib}

\begin{thebibliography}{}

\bibitem[Ali et~al., 2013]{ali_achievement_2013}
Ali, M.~K., Bullard, K.~M., Saaddine, J.~B., Cowie, C.~C., Imperatore, G., and
  Gregg, E.~W. (2013).
\newblock Achievement of goals in {U}.{S}. diabetes care, 1999-2010.
\newblock {\em The New England Journal of Medicine}, 368(17):1613--1624.

\bibitem[Austin, 2009]{austin2009balance}
Austin, P.~C. (2009).
\newblock Balance diagnostics for comparing the distribution of baseline
  covariates between treatment groups in propensity-score matched samples.
\newblock {\em Statistics in Medicine}, 28(25):3083--3107.

\bibitem[Bareinboim and Pearl, 2016]{bareinboim_causal_2016}
Bareinboim, E. and Pearl, J. (2016).
\newblock Causal inference and the data-fusion problem.
\newblock {\em Proceedings of the National Academy of Sciences},
  113(27):7345--7352.

\bibitem[Berkowitz et~al., 2014]{berkowitz_initial_2014}
Berkowitz, S.~A., Krumme, A.~A., Avorn, J., Brennan, T., Matlin, O.~S.,
  Spettell, C.~M., Pezalla, E.~J., Brill, G., Shrank, W.~H., and Choudhry,
  N.~K. (2014).
\newblock Initial choice of oral glucose-lowering medication for diabetes
  mellitus: a patient-centered comparative effectiveness study.
\newblock {\em JAMA internal medicine}, 174(12):1955--1962.

\bibitem[Chan et~al., 2016]{chan2016globally}
Chan, K. C.~G., Yam, S. C.~P., and Zhang, Z. (2016).
\newblock Globally efficient non-parametric inference of average treatment
  effects by empirical balancing calibration weighting.
\newblock {\em Journal of the Royal Statistical Society. Series B, Statistical
  methodology}, 78(3):673.

\bibitem[Cheng et~al., 2018]{cheng_trends_2018}
Cheng, Y.~J., Imperatore, G., Geiss, L.~S., Saydah, S.~H., Albright, A.~L.,
  Ali, M.~K., and Gregg, E.~W. (2018).
\newblock Trends and {disparities} in {cardiovascular} {mortality} {among}
  {U}.{S}. {adults} {with} and {without} {self}-{reported} {diabetes},
  1988-2015.
\newblock {\em Diabetes Care}, 41(11):2306--2315.

\bibitem[Dahabreh et~al., 2020]{dahabreh2020double}
Dahabreh, I.~J., Robertson, S.~E., Steingrimsson, J.~A., Stuart, E.~A., and
  Hernán, M.~A. (2020).
\newblock Extending inferences from a randomized trial to a new target
  population.
\newblock {\em Statistics in Medicine}, 39(14):1999--2014.

\bibitem[Desai et~al., 2012]{desai_patterns_2012}
Desai, N.~R., Shrank, W.~H., Fischer, M.~A., Avorn, J., Liberman, J.~N.,
  Schneeweiss, S., Pakes, J., Brennan, T.~A., and Choudhry, N.~K. (2012).
\newblock Patterns of medication initiation in newly diagnosed diabetes
  mellitus: quality and cost implications.
\newblock {\em The American Journal of Medicine}, 125(3):302.e1--7.

\bibitem[Deville and S{\"a}rndal, 1992]{deville_calibration_1992}
Deville, J.-C. and S{\"a}rndal, C.-E. (1992).
\newblock Calibration {estimators} in {survey} {sampling}.
\newblock {\em Journal of the American Statistical Association},
  87(418):376--382.

\bibitem[Fan et~al., 2021]{fan_improving_2018}
Fan, J., Imai, K., Lee, I., Liu, H., Ning, Y., and Yang, X. (2021).
\newblock Optimal covariate balancing conditions in propensity score
  estimation.
\newblock {\em Journal of Business \& Economic Statistics}, pages 1--14.

\bibitem[Geiss et~al., 2014]{geiss_prevalence_2014}
Geiss, L.~S., Wang, J., Cheng, Y.~J., Thompson, T.~J., Barker, L., Li, Y.,
  Albright, A.~L., and Gregg, E.~W. (2014).
\newblock Prevalence and incidence trends for diagnosed diabetes among adults
  aged 20 to 79 years, {United} {States}, 1980-2012.
\newblock {\em JAMA}, 312(12):1218--1226.

\bibitem[Geman and Hwang, 1982]{geman_nonparametric_1982}
Geman, S. and Hwang, C.-R. (1982).
\newblock Nonparametric {maximum} {likelihood} {estimation} by the {method} of
  {sieves}.
\newblock {\em The Annals of Statistics}, 10(2):401--414.

\bibitem[Gregg et~al., 2018]{gregg_trends_2018}
Gregg, E.~W., Cheng, Y.~J., Srinivasan, M., Lin, J., Geiss, L.~S., Albright,
  A.~L., and Imperatore, G. (2018).
\newblock Trends in cause-specific mortality among adults with and without
  diagnosed diabetes in the {USA}: an epidemiological analysis of linked
  national survey and vital statistics data.
\newblock {\em Lancet (London, England)}, 391(10138):2430--2440.

\bibitem[Gregg et~al., 2014]{gregg_changes_2014}
Gregg, E.~W., Li, Y., Wang, J., Burrows, N.~R., Ali, M.~K., Rolka, D.,
  Williams, D.~E., and Geiss, L. (2014).
\newblock Changes in diabetes-related complications in the {United} {States},
  1990-2010.
\newblock {\em The New England Journal of Medicine}, 370(16):1514--1523.

\bibitem[Gruber and {van der Laan}, 2010]{gruber2010targeted}
Gruber, S. and {van der Laan}, M.~J. (2010).
\newblock A targeted maximum likelihood estimator of a causal effect on a
  bounded continuous outcome.
\newblock {\em The International Journal of Biostatistics}, 6(1).

\bibitem[Hampp et~al., 2014]{hampp_use_2014}
Hampp, C., Borders-Hemphill, V., Moeny, D.~G., and Wysowski, D.~K. (2014).
\newblock Use of antidiabetic drugs in the {U}.{S}., 2003-2012.
\newblock {\em Diabetes Care}, 37(5):1367--1374.

\bibitem[Hirshberg et~al., 2019]{hirshberg2019minimax}
Hirshberg, D.~A., Maleki, A., and Zubizarreta, J.~R. (2019).
\newblock Minimax linear estimation of the retargeted mean.
\newblock {\em arXiv preprint arXiv:1901.10296}.

\bibitem[Holman et~al., 2017]{holman_effects_2017}
Holman, R.~R., Bethel, M.~A., Mentz, R.~J., Thompson, V.~P., Lokhnygina, Y.,
  Buse, J.~B., Chan, J.~C., Choi, J., Gustavson, S.~M., Iqbal, N., Maggioni,
  A.~P., Marso, S.~P., {\"O}hman, P., Pagidipati, N.~J., Poulter, N.,
  Ramachandran, A., Zinman, B., and Hernandez, A.~F. (2017).
\newblock Effects of {once}-{weekly} {exenatide} on {cardiovascular} {outcomes}
  in {type} 2 {diabetes}.
\newblock {\em New England Journal of Medicine}, 377(13):1228--1239.

\bibitem[Hong et~al., 2013]{hong_effects_2013}
Hong, J., Zhang, Y., Lai, S., Lv, A., Su, Q., Dong, Y., Zhou, Z., Tang, W.,
  Zhao, J., Cui, L., Zou, D., Wang, D., Li, H., Liu, C., Wu, G., Shen, J., Zhu,
  D., Wang, W., Shen, W., Ning, G., and {SPREAD-DIMCAD Investigators} (2013).
\newblock Effects of metformin versus glipizide on cardiovascular outcomes in
  patients with type 2 diabetes and coronary artery disease.
\newblock {\em Diabetes Care}, 36(5):1304--1311.

\bibitem[Josey et~al., 2021a]{josey2021transporting}
Josey, K.~P., Berkowitz, S.~A., Ghosh, D., and Raghavan, S. (2021a).
\newblock Transporting experimental results with entropy balancing.
\newblock {\em Statistics in Medicine}, 40(19):4310--4326.

\bibitem[Josey et~al., 2021b]{josey2021framework}
Josey, K.~P., Juarez-Colunga, E., Yang, F., and Ghosh, D. (2021b).
\newblock A framework for covariate balance using bregman distances.
\newblock {\em Scandinavian Journal of Statistics}, 48(3):790--816.

\bibitem[Kang and Schafer, 2007]{kang_demystifying_2007}
Kang, J. D.~Y. and Schafer, J.~L. (2007).
\newblock Demystifying {double} {robustness}: {a} {comparison} of {alternative}
  {strategies} for {estimating} a {population} {mean} from {incomplete} {data}.
\newblock {\em Statistical Science}, 22(4):523--539.

\bibitem[Keele et~al., 2020]{keele2020hospital}
Keele, L., Ben-Michael, E., Feller, A., Kelz, R., and Miratrix, L. (2020).
\newblock Hospital quality risk standardization via approximate balancing
  weights.
\newblock {\em arXiv preprint arXiv:2007.09056}.

\bibitem[Lee et~al., 2021]{lee2021improving}
Lee, D., Yang, S., Dong, L., Wang, X., Zeng, D., and Cai, J. (2021).
\newblock Improving trial generalizability using observational studies.
\newblock {\em Biometrics}.

\bibitem[Lesko et~al., 2017]{lesko_generalizing_2017}
Lesko, C.~R., Buchanan, A.~L., Westreich, D., Edwards, J.~K., Hudgens, M.~G.,
  and Cole, S.~R. (2017).
\newblock Generalizing study results: a potential outcomes perspective.
\newblock {\em Epidemiology (Cambridge, Mass.)}, 28(4):553--561.

\bibitem[Lu et~al., 2021]{lu2021you}
Lu, B., Ben-Michael, E., Feller, A., and Miratrix, L. (2021).
\newblock Is it who you are or where you are? accounting for compositional
  differences in cross-site treatment variation.
\newblock {\em arXiv preprint arXiv:2103.14765}.

\bibitem[Marso et~al., 2016a]{marso_semaglutide_2016}
Marso, S.~P., Bain, S.~C., Consoli, A., Eliaschewitz, F.~G., J{\'o}dar, E.,
  Leiter, L.~A., Lingvay, I., Rosenstock, J., Seufert, J., Warren, M.~L., Woo,
  V., Hansen, O., Holst, A.~G., Pettersson, J., and Vilsb{\o}ll, T. (2016a).
\newblock Semaglutide and {cardiovascular} {outcomes} in {patients} with {type}
  2 {diabetes}.
\newblock {\em New England Journal of Medicine}, 375(19):1834--1844.

\bibitem[Marso et~al., 2016b]{marso_liraglutide_2016}
Marso, S.~P., Daniels, G.~H., Brown-Frandsen, K., Kristensen, P., Mann, J.~F.,
  Nauck, M.~A., Nissen, S.~E., Pocock, S., Poulter, N.~R., Ravn, L.~S.,
  Steinberg, W.~M., Stockner, M., Zinman, B., Bergenstal, R.~M., and Buse,
  J.~B. (2016b).
\newblock Liraglutide and {cardiovascular} {outcomes} in {type} 2 {diabetes}.
\newblock {\em New England Journal of Medicine}, 375(4):311--322.

\bibitem[Neal et~al., 2017]{neal_canagliflozin_2017}
Neal, B., Perkovic, V., Mahaffey, K.~W., de~Zeeuw, D., Fulcher, G., Erondu, N.,
  Shaw, W., Law, G., Desai, M., and Matthews, D.~R. (2017).
\newblock Canagliflozin and {cardiovascular} and {renal} {events} in {type} 2
  {diabetes}.
\newblock {\em New England Journal of Medicine}, 377(7):644--657.

\bibitem[Pearl and Bareinboim, 2014]{pearl_external_2014}
Pearl, J. and Bareinboim, E. (2014).
\newblock External {validity}: {from} {do}-{calculus} to {transportability}
  {across} {populations}.
\newblock {\em Statistical Science}, 29(4):579--595.

\bibitem[Raghavan et~al., 2019]{raghavan_diabetes_2019}
Raghavan, S., Vassy, J.~L., Ho, Y.-L., Song, R.~J., Gagnon, D.~R., Cho, K.,
  Wilson, P. W.~F., and Phillips, L.~S. (2019).
\newblock Diabetes {mellitus}-{related} {all}-{cause} and {cardiovascular}
  {mortality} in a {national} {cohort} of {adults}.
\newblock {\em Journal of the American Heart Association}, 8(4):e011295.

\bibitem[Rao Kondapally~Seshasai et~al.,
  2011]{rao_kondapally_seshasai_diabetes_2011}
Rao Kondapally~Seshasai, S., Kaptoge, S., Thompson, A., Di~Angelantonio, E.,
  Gao, P., Sarwar, N., Whincup, P.~H., Mukamal, K.~J., Gillum, R.~F., Holme,
  I., Nj{\o}lstad, I., Fletcher, A., Nilsson, P., Lewington, S., Collins, R.,
  Gudnason, V., Thompson, S.~G., Sattar, N., Selvin, E., Hu, F.~B., Danesh, J.,
  and {Emerging Risk Factors Collaboration} (2011).
\newblock Diabetes mellitus, fasting glucose, and risk of cause-specific death.
\newblock {\em The New England Journal of Medicine}, 364(9):829--841.

\bibitem[Robins et~al., 1994]{robins_estimation_1994}
Robins, J.~M., Rotnitzky, A., and Zhao, L.~P. (1994).
\newblock Estimation of {regression} {coefficients} {when} {some} {regressors}
  are not {always} {observed}.
\newblock {\em Journal of the American Statistical Association}, 89:846--866.

\bibitem[Rosenbaum and Rubin, 1983]{rosenbaum_central_1983}
Rosenbaum, P.~R. and Rubin, D.~B. (1983).
\newblock The {central} {role} of the {propensity} {score} in {observational}
  {studies} for {causal} {effects}.
\newblock {\em Biometrika}, 70(1):41--55.

\bibitem[Roumie et~al., 2012]{roumie_comparative_2012}
Roumie, C.~L., Hung, A.~M., Greevy, R.~A., Grijalva, C.~G., Liu, X., Murff,
  H.~J., Elasy, T.~A., and Griffin, M.~R. (2012).
\newblock Comparative effectiveness of sulfonylurea and metformin monotherapy
  on cardiovascular events in type 2 diabetes mellitus: a cohort study.
\newblock {\em Annals of Internal Medicine}, 157(9):601--610.

\bibitem[Rubin, 1974]{rubin_estimating_1974}
Rubin, D.~B. (1974).
\newblock Estimating causal effects of treatments in randomized and
  nonrandomized studies.
\newblock {\em Journal of Educational Psychology}, 66(5):688--701.

\bibitem[Rudolph and {van der Laan}, 2017]{rudolph_robust_2017}
Rudolph, K.~E. and {van der Laan}, M.~J. (2017).
\newblock Robust estimation of encouragement design intervention effects
  transported across sites.
\newblock {\em Journal of the Royal Statistical Society: Series B (Statistical
  Methodology)}, 79(5):1509--1525.

\bibitem[Schramm et~al., 2011]{schramm_mortality_2011}
Schramm, T.~K., Gislason, G.~H., Vaag, A., Rasmussen, J.~N., Folke, F., Hansen,
  M.~L., Fosb{\o}l, E.~L., K{\o}ber, L., Norgaard, M.~L., Madsen, M., Hansen,
  P.~R., and Torp-Pedersen, C. (2011).
\newblock Mortality and cardiovascular risk associated with different insulin
  secretagogues compared with metformin in type 2 diabetes, with or without a
  previous myocardial infarction: a nationwide study.
\newblock {\em European Heart Journal}, 32(15).

\bibitem[Schuler and Rose, 2017]{schuler_targeted_2017}
Schuler, M.~S. and Rose, S. (2017).
\newblock Targeted {maximum} {likelihood} {estimation} for {causal} {inference}
  in {observational} {studies}.
\newblock {\em American Journal of Epidemiology}, 185(1):65--73.

\bibitem[Selvin et~al., 2014]{selvin_trends_2014}
Selvin, E., Parrinello, C.~M., Sacks, D.~B., and Coresh, J. (2014).
\newblock Trends in prevalence and control of diabetes in the {United}
  {States}, 1988-1994 and 1999-2010.
\newblock {\em Annals of Internal Medicine}, 160(8):517--525.

\bibitem[Signorovitch et~al., 2010]{signorovitch_comparative_2010}
Signorovitch, J.~E., Wu, E.~Q., Yu, A.~P., Gerrits, C.~M., Kantor, E., Bao, Y.,
  Gupta, S.~R., and Mulani, P.~M. (2010).
\newblock Comparative {effectiveness} {without} {head}-to-{head} {trials}.
\newblock {\em PharmacoEconomics}, 28(10):935--945.

\bibitem[Stefanski and Boos, 2002]{stefanski_calculus_2002}
Stefanski, L.~A. and Boos, D.~D. (2002).
\newblock The {calculus} of {M}-{estimation}.
\newblock {\em The American Statistician}, 56(1):29--38.

\bibitem[{The BARI 2D Study Group},
  2009]{the_bari_2d_study_group_randomized_2009}
{The BARI 2D Study Group} (2009).
\newblock A {Randomized} {Trial} of {Therapies} for {Type} 2 {Diabetes} and
  {Coronary} {Artery} {Disease}.
\newblock {\em New England Journal of Medicine}, 360(24):2503--2515.

\bibitem[Tsiatis, 2006]{tsiatis_semiparametric_2006}
Tsiatis, A. (2006).
\newblock {\em Semiparametric {Theory} and {Missing} {Data}}.
\newblock Springer {Series} in {Statistics}. Springer-Verlag, New York.

\bibitem[{van der Laan} and Rubin, 2006]{laan_targeted_2006}
{van der Laan}, M.~J. and Rubin, D. (2006).
\newblock Targeted {maximum} {likelihood} {learning}.
\newblock {\em The International Journal of Biostatistics}, 2(1).

\bibitem[VanderWeele and Shpitser, 2013]{vanderweele2013definition}
VanderWeele, T.~J. and Shpitser, I. (2013).
\newblock On the definition of a confounder.
\newblock {\em Annals of Statistics}, 41(1):196--220.

\bibitem[Varvaki~Rados et~al., 2016]{varvaki_rados_association_2016}
Varvaki~Rados, D., Catani~Pinto, L., Reck~Remonti, L., Bauermann~Leit{\~a}o,
  C., and Gross, J.~L. (2016).
\newblock The {association} between {sulfonylurea} {use} and {all}-{cause} and
  {cardiovascular} {mortality}: {a} {meta}-{analysis} with {trial} {sequential}
  {analysis} of {randomized} {clinical} {trials}.
\newblock {\em PLoS medicine}, 13(4):e1001992.

\bibitem[Wang and Zubizarreta, 2020]{wang2020minimal}
Wang, Y. and Zubizarreta, J.~R. (2020).
\newblock Minimal dispersion approximately balancing weights: asymptotic
  properties and practical considerations.
\newblock {\em Biometrika}, 107(1):93--105.

\bibitem[Westreich et~al., 2017]{westreich_transportability_2017}
Westreich, D., Edwards, J.~K., Lesko, C.~R., Stuart, E., and Cole, S.~R.
  (2017).
\newblock Transportability of {trial} {results} {using} {inverse} {odds} of
  {sampling} {weights}.
\newblock {\em American Journal of Epidemiology}, 186(8):1010--1014.

\bibitem[Wheeler et~al., 2013]{wheeler_mortality_2013}
Wheeler, S., Moore, K., Forsberg, C.~W., Riley, K., Floyd, J.~S., Smith, N.~L.,
  and Boyko, E.~J. (2013).
\newblock Mortality among veterans with type 2 diabetes initiating metformin,
  sulfonylurea or rosiglitazone monotherapy.
\newblock {\em Diabetologia}, 56(9):1934--1943.

\bibitem[Zhao and Percival, 2017]{zhao_entropy_2017}
Zhao, Q. and Percival, D. (2017).
\newblock Entropy {balancing} is {doubly} {robust}.
\newblock {\em Journal of Causal Inference}, 5.

\bibitem[Zinman et~al., 2015]{zinman_empagliflozin_2015}
Zinman, B., Wanner, C., Lachin, J.~M., Fitchett, D., Bluhmki, E., Hantel, S.,
  Mattheus, M., Devins, T., Johansen, O.~E., Woerle, H.~J., Broedl, U.~C., and
  Inzucchi, S.~E. (2015).
\newblock Empagliflozin, {cardiovascular} {outcomes}, and {mortality} in {type}
  2 {diabetes}.
\newblock {\em New England Journal of Medicine}, 373(22):2117--2128.

\end{thebibliography}

\newpage

\appendix

\section{Iteratively Updated Calibration Weights}

The primal/dual problems presented in the main article can be estimated in an iterative fashion. A quick breakdown of the sequential procedure for the transportability problem is as follows. First, we balance the study and target sample covariate moments by estimating the sampling weights. Second, we estimate balancing weights across the within the study samples given these estimated sampling weights from the previous step. We iterate between finding the sampling and balancing weights until convergence is achieved. After iterating between estimating these two weights, we estimate the target population average treatment effect using a Hajek-type estimator.

To begin, we initialize $\hat{p}'(S_i,\mathbf{X}_i,Z_i) = 1$ for all $i = 1,2\ldots, n$. The first step of estimating the sampling weights is solved in relatively the same manner as in (6) and (7) - solving for \begin{equation}\label{dual-base:04}
\hat{\boldsymbol{\gamma}}' = \argmax_{\boldsymbol{\gamma} \in \Re^{m}} \ \sum_{\{i:S_i = 1\}} \left\{ -\hat{p}'(S_i,\mathbf{X}_i,Z_i)\exp\left[-S_i\sum_{j = 1}^m  c_j(\mathbf{X}_i) \gamma_j\right] - \sum_{j = 1}^m \hat{\theta}_{0j} \gamma_j \right\}.
\end{equation}
With $\hat{\boldsymbol{\gamma}}' = (\hat{\gamma}'_1, \hat{\gamma}'_2, \ldots, \hat{\gamma}'_m)^{T}$, we update the sampling weights with
\begin{equation}\label{weights-base:04}
\hat{q}'(S_i,\mathbf{X}_i,Z_i) = \hat{p}'(S_i,\mathbf{X}_i,Z_i)\exp\left[-S_i\sum_{j = 1}^m c_j(\mathbf{X}_i)\hat{\gamma}'_j \right].
\end{equation}

The second step is to compute balancing weights that mitigate treatment group heterogeneity by solving for
\begin{equation}\label{dual-cbps:04}
\hat{\boldsymbol{\lambda}}' = \argmax_{\boldsymbol{\lambda} \in \Re^{m}} \ \sum_{\{i:S_i = 1\}}  - \hat{q}'(S_i,\mathbf{X}_i,Z_i)\exp\left[-(2Z_i - 1) \sum_{j = 1}^m  c_j(\mathbf{X}_i)\lambda_{j} \right]
\end{equation} 
where $\hat{\boldsymbol{\lambda}}' \equiv (\hat{\lambda}'_{1}, \hat{\lambda}'_{2}, \ldots, \hat{\lambda}'_{m})^T$. The resulting balancing weights are estimated with 
\begin{equation}\label{weights-cbps:04}
\hat{p}'(S_i,\mathbf{X}_i,Z_i) = \hat{q}'(S_i,\mathbf{X}_i,Z_i) \exp\left[-S_i(2Z_i - 1)\sum_{j = 1}^m  c_j(\mathbf{X}_i)\hat{\lambda}'_{j} \right]
\end{equation} 
We then iterate between updating $\hat{q}'(S_i,\mathbf{X}_i,Z_i)$ in (\ref{dual-base:04})-(\ref{weights-base:04}) and $\hat{p}'(S_i,\mathbf{X}_i,Z_i)$ in (\ref{dual-cbps:04})-(\ref{weights-cbps:04}). After several iterations, the weights from (\ref{weights-cbps:04}) should converge and be identical to the balancing weights presented in Section 3.3 -- $\hat{p}'(S_i,\mathbf{X}_i,Z_i) \rightarrow \hat{p}(S_i,\mathbf{X}_i,Z_i)$. This version to the weighting estimator used by the full calibration approach better explains how the separate sampling and treatment weights interact with one another. To estimate the target population average treatment effect, we use a Hajek-type estimator, which solves for
\[ \hat{\tau}'_{\text{CAL}} = \frac{1}{n_1}\sum_{\{i: S_i = 1\}} \hat{p}'(S_i,\mathbf{X}_i,Z_i)(2Z_i - 1)Y_i. \]

\section{Inference for Data-Fusion using Calibration Weights }\label{proof:04}

The following proof is largely inspired by \cite{zhao_entropy_2017}. Consider the full calibration data-fusion estimator from Section 4, but under the equivalent primal problem to (13), which solves
\begin{equation}\label{primal-fusion-2:04}
\begin{split} 
\text{minimize} &\enskip \sum_{i = 1}^n \left\{p(S_i, \mathbf{X}_i, Z_i) \log\left[p(S_i, \mathbf{X}_i, Z_i)\right] - p(S_i, \mathbf{X}_i, Z_i) \right\} \\
\text{subject to} &\enskip \sum_{i = 1}^n  S_i Z_i p(S_i, \mathbf{X}_i, Z_i) c_j(\mathbf{X}_i) = \frac{n_1\hat{\theta}_{0j}}{2}, \\
&\enskip \sum_{i = 1}^n  S_i p(S_i, \mathbf{X}_i, Z_i) c_j(\mathbf{X}_i) = n_1\hat{\theta}_{0j}, \\
&\enskip \sum_{i = 1}^n  (1 - S_i)Z_i p(S_i, \mathbf{X}_i, Z_i) c_j(\mathbf{X}_i) = \frac{n_0\hat{\theta}_{0j}}{2} \enskip \text{and} \\
&\enskip \sum_{i = 1}^n  (1 - S_i) p(S_i, \mathbf{X}_i, Z_i) c_j(\mathbf{X}_i) = n_0 \hat{\theta}_{0j} \enskip \text{for all} \enskip j = 1,2,\ldots,m.
\end{split}
\end{equation} 
We begin by defining the estimating equations for the parameters of the full calibration approach subject to (\ref{primal-fusion-2:04}). To help simplify the notation, let $\mathbf{c}(\mathbf{X}_i) \equiv [c_1(\mathbf{X}_i), c_2(\mathbf{X}_i), \ldots, c_m(\mathbf{X}_i)]^T$. We define $\boldsymbol{\omega}(S_i, \mathbf{X}_i; \boldsymbol{\theta}_0) \equiv (1 - S_i)\left[ \mathbf{c}(\mathbf{X}_i) - \boldsymbol{\theta}_0\right]$, which is solved by $\sum_{i = 1}^n \boldsymbol{\omega}(S_i, \mathbf{X}_i; \hat{\boldsymbol{\theta}}_0)] = \mathbf{0}_m$ where $\hat{\boldsymbol{\theta}}_0 = \frac{1}{n_0} \sum_{i = 1}^n (1 - S_i)\mathbf{c}(\mathbf{X}_i)$. Next, define 
\[\begin{split} \boldsymbol{\zeta}_0(S_i, \mathbf{X}_i; \boldsymbol{\gamma}_0, \boldsymbol{\theta}_0) &\equiv (1 - S_i)\exp\left[-\sum_{j = 1}^m c_j(\mathbf{X}_i)\gamma_{0j}\right]\left[\mathbf{c}(\mathbf{X}_i) - \boldsymbol{\theta}_0\right] \enskip \text{and} \\
\boldsymbol{\zeta}_1(S_i, \mathbf{X}_i; \boldsymbol{\gamma}_1, \boldsymbol{\theta}_0) &\equiv S_i\exp\left[-\sum_{j = 1}^m c_j(\mathbf{X}_i)\gamma_{1j}\right]\left[\mathbf{c}(\mathbf{X}_i) - \boldsymbol{\theta}_0\right]
\end{split}\] which is used to solve for
$\sum_{i = 1}^n \zeta_0(S_i, \mathbf{X}_i; \tilde{\boldsymbol{\gamma}}_0, \hat{\boldsymbol{\theta}}_0) = \mathbf{0}_m$ and $\sum_{i = 1}^n \zeta_1(S_i, \mathbf{X}_i; \tilde{\boldsymbol{\gamma}}_1, \hat{\boldsymbol{\theta}}_0) = \mathbf{0}_m$ as a result of the Lagrangian multiplier theorem. The score equations for $\boldsymbol{\delta}_0$ and $\boldsymbol{\delta}_1$, replacing $\boldsymbol{\lambda}_0$ and $\boldsymbol{\lambda}_1$ after changing the primal problem to (\ref{primal-fusion-2:04}), are identified as
\[\begin{split}
\boldsymbol{\xi}_0(S_i, \mathbf{X}_i, Z_i; \boldsymbol{\gamma}_0, \boldsymbol{\delta}_0, \boldsymbol{\theta}_0) &\equiv (1 - S_i)Z_i\exp\left[-\sum_{j = 1}^m c_j(\mathbf{X}_i)(\gamma_{0j} + \delta_{0j})\right]\left[\mathbf{c}(\mathbf{X}_i) - \boldsymbol{\theta}_0\right] \enskip \text{and} \\
\boldsymbol{\xi}_1(S_i, \mathbf{X}_i, Z_i; \boldsymbol{\gamma}_1, \boldsymbol{\delta}_1, \boldsymbol{\theta}_0) &\equiv S_i Z_i\exp\left[-\sum_{j = 1}^m c_j(\mathbf{X}_i)(\gamma_{1j} + \delta_{1j})\right] \left[\mathbf{c}(\mathbf{X}_i) - \boldsymbol{\theta}_0\right].
\end{split} \]
which can be used to solve for $\sum_{i = 1}^n \boldsymbol{\xi}_0(S_i, \mathbf{X}_i, Z_i; \tilde{\boldsymbol{\gamma}}_0, \tilde{\boldsymbol{\delta}}_0, \hat{\boldsymbol{\theta}}_0) = \mathbf{0}_m$ and $\sum_{i = 1}^n \boldsymbol{\xi}_1(S_i, \mathbf{X}_i, Z_i; \tilde{\boldsymbol{\gamma}}_1, \tilde{\boldsymbol{\delta}}_1, \hat{\boldsymbol{\theta}}_0) = \mathbf{0}_m$ once first finding $\tilde{\boldsymbol{\gamma}}_0$ and $\tilde{\boldsymbol{\gamma}}_1$. Finally, we can write the score equation for $\tau_0$ as 
\begin{equation}\label{phi:04}
\begin{split}
\phi(\mathbf{X}_i, Y_i, Z_i; \boldsymbol{\gamma}_0, \boldsymbol{\gamma}_1, \boldsymbol{\delta}_0, \boldsymbol{\delta}_1, \tau_0) &\equiv S_i\Bigg\{Z_i\exp\left[-\sum_{j = 1}^m c_j(\mathbf{X}_i)(\gamma_{1j} + \delta_{1j})\right]\left[Y_i(1) - \tau_0\right] \\
&\qquad - (1 - Z_i)\exp\left[-\sum_{j = 1}^m c_j(\mathbf{X}_i)\gamma_{1j}\right]Y_i(0)\Bigg\} \\
&\quad + (1 - S_i)\Bigg\{Z_i\exp\left[-\sum_{j = 1}^m c_j(\mathbf{X}_i)(\gamma_{0j} + \delta_{0j})\right]\left[Y_i(1) - \tau_0\right] \\
&\qquad - (1 - Z_i)\exp\left[-\sum_{j = 1}^m c_j(\mathbf{X}_i)\gamma_{0j}\right]Y_i(0)\Bigg\}.
\end{split}
\end{equation} which we may solve as $\sum_{i = 1}^n \phi(S_i, \mathbf{X}_i, Y_i Z_i; \tilde{\boldsymbol{\gamma}}_0, \tilde{\boldsymbol{\gamma}}_1, \tilde{\boldsymbol{\delta}}_0, \tilde{\boldsymbol{\delta}}_1, \tilde{\tau}_{\text{CAL}}) = 0$. Given this setup, we can see that these equations are conducive of M-estimation theory. For simplicity, we will often drop the parameter values in the notation when using the functional representations of the estimating equations.

To show double-robustness, we first prove that $\tilde{\tau}_{\text{CAL}}$ is consistent for $\tau_0$ given Assumption 5. This means we can assume
 \[\begin{split}
    \mu_0(S_i, \mathbf{X}_i) &= (1 - S_i)\sum_{j = 1}^{m} c_j(\mathbf{X}_i)\beta_{0j} + S_i\sum_{j = 1}^{m} c_j(\mathbf{X}_i)\beta_{1j} \enskip \text{and} \\
    \mu_1(S_i, \mathbf{X}_i)  &= \mu_0(S_i, \mathbf{X}_i) + \sum_{j = 1}^{m} c_j(\mathbf{X}_i)\alpha_j
 \end{split} \] 
 Let $\tilde{p}(S_i, \mathbf{X}_i, Z_i)$ be determined by (15) where $\tilde{\boldsymbol{\lambda}}_0$, $\tilde{\boldsymbol{\lambda}}_1$, $\tilde{\boldsymbol{\gamma}}_0$, and $\tilde{\boldsymbol{\gamma}}_1$ are solved using the objective functions in (14). If we assume $c_1(\mathbf{X}_i) = 1$ for all $i = \{i:S_i = 1\}$, then $n_1 = \sum_{\{i:S_i = 1\}} \hat{p}(S_i,\mathbf{X}_i,Z_i)Z_i$. If we substitute $\tilde{p}(S_i, \mathbf{X}_i, Z_i)$ into (\ref{phi:04}), regardless of whether it is a correctly specified model for the balancing and sampling weights, we get
\[ \begin{split}
\mathbb{E}\left[\sum_{i = 1}^n \phi(S_i, \mathbf{X}_i, Y_i, Z_i)\right] 
&= \mathbb{E}\left\{\sum_{i = 1}^n \mathbb{E}\left[Z_i\tilde{p}(S_i,\mathbf{X}_i, Z_i)Y_i(1) - (1 - Z_i)\tilde{p}(S_i, \mathbf{X}_i, Z_i)Y_i(0) | \mathbf{X}_i, Z_i\right]\right\} - \tau_0 \\
&= \mathbb{E}\left[\sum_{i = 1}^n Z_i\tilde{p}(S_i, \mathbf{X}_i, Z_i)\mu_1(S_i, \mathbf{X}_i) - \sum_{i = 1}^n (1 - Z_i)\tilde{p}(S_i, \mathbf{X}_i, Z_i)\mu_0(S_i, \mathbf{X}_i)\right] - \tau_0 \\
&= \mathbb{E}\Bigg\{\sum_{i = 1}^n Z_i\tilde{p}(S_i, \mathbf{X}_i, Z_i)\left[\mu_0(S_i, \mathbf{X}_i) + \sum_{j = 1}^{m} c_j(\mathbf{X}_i)\alpha_j\right] \\
&\qquad - \sum_{i = 1}^n (1 - Z_i)\tilde{p}(S_i,\mathbf{X}_i,Z_i)\mu_0(S_i, \mathbf{X}_i)\Bigg\} - \tau_0 \\
&= \mathbb{E}\left\{\sum_{i = 1}^n (2Z_i - 1)\tilde{p}(S_i,\mathbf{X}_i,Z_i)\mu_0(S_i, \mathbf{X}_i) + \sum_{i = 1}^n Z_i\tilde{p}(\mathbf{X}_i)\left[\sum_{j = 1}^{m} c_j(\mathbf{X}_i)\alpha_j\right]\right\} - \tau_0 \\
&= \mathbb{E}\Bigg[\sum_{\{i:S_i = 0\}} (2Z_i - 1)\tilde{p}(S_i,\mathbf{X}_i,Z_i) \sum_{j = 1}^m c_j(\mathbf{X}_i)\beta_{0j}  \\
&\qquad + \sum_{\{i:S_i = 1\}} (2Z_i - 1)\tilde{p}(S_i,\mathbf{X}_i,Z_i) \sum_{j = 1}^m c_{j}(\mathbf{X}_i)\beta_{1j} + \sum_{j = 1}^m \hat{\theta}_{0j}\alpha_j \Bigg] - \tau_0 \\
&= \mathbb{E}\Bigg[\sum_{j = 1}^m \hat{\theta}_{0j}\alpha_j \Bigg] - \tau_0 = 0
\end{split} \]

Now suppose Assumptions 6 and 7 are given. This means 
\[ \begin{split}
\text{logit}[\pi_s(\mathbf{X}_i)] &= S_i\sum_{j = 1}^{m} c_j(\mathbf{X}_i)\lambda^{\dag}_{1j} + (1 - S_i)\sum_{j = 1}^{m} c_j(\mathbf{X}_i)\lambda^{\dag}_{0j} \enskip \text{and} \\
\text{logit}[\rho(\mathbf{X}_i)] &= \sum_{j = 1}^{m} c_j(\mathbf{X}_i)\gamma^{\dag}_j. 
\end{split} \] An equivalent representation for the probability of $Z_i$ conditioned on $\mathbf{X}_i$ and $S_i = 0$ is \[ \pi_0(\mathbf{X}_i) = \frac{\exp\left[\sum_{j = 1}^{m} c_j(\mathbf{X}_i)\delta^{\dag}_{1j}\right]}{\exp\left[\sum_{j = 1}^{m} c_j(\mathbf{X}_i)\delta^{\dag}_{1j}\right] + \exp\left[\sum_{j = 1}^{m} c_j(\mathbf{X}_i)\delta^{\dag}_{0j}\right]} \] for some $\delta^{\dag}_{1j}, \delta^{\dag}_{0j} \in \Re$ and $j = 1,2,\ldots, m$. It is trivial to see that $\mathbb{E}[\boldsymbol{\omega}(S_i, \mathbf{X}_i; \boldsymbol{\theta}_0)] = 0$, where $\boldsymbol{\theta}_0 \equiv \mathbb{E}[\mathbf{c}(\mathbf{X}_i)|S_i = 0]$. The expectation of the estimating equation for $\boldsymbol{\gamma}_0$ can be expanded into
\[\begin{split} 
\mathbb{E}\left[\boldsymbol{\zeta}_0(S_i, \mathbf{X}_i; \boldsymbol{\gamma}_0, \boldsymbol{\theta}_0)\right] &= \mathbb{E}\left(\mathbb{E}\left\{(1 - S_i)\exp\left[-\sum_{j = 1}^m c_j(\mathbf{X}_i)\gamma_{0j}\right]\left[\mathbf{c}(\mathbf{X}_i) - \boldsymbol{\theta}_0\right]\middle|\mathbf{X}_i\right\}\right) \\
&= \mathbb{E}\left\{\frac{\exp\left[-\sum_{j = 1}^m c_j(\mathbf{X}_i)\gamma_{0j}\right]}{1 + \exp\left[\sum_{j = 1}^m c_j(\mathbf{X}_i)\gamma^{\dag}_j\right]}\left[\mathbf{c}(\mathbf{X}_i) - \boldsymbol{\theta}_0\right]\right\},
\end{split}\]
which can only evaluate to zero if \[ \frac{\exp\left[-\sum_{j = 1}^m c_j(\mathbf{X}_i)\gamma_{0j}\right]}{1 + \exp\left[\sum_{j = 1}^m c_j(\mathbf{X}_i)\gamma^{\dag}_j\right]} \propto \Pr\{S_i = 0|\mathbf{X}_i\} \] so that $ \mathbb{E}\left[\boldsymbol{\zeta}(S_i, \mathbf{X}_i; \boldsymbol{\gamma}, \boldsymbol{\theta}_0)\right] \propto \mathbb{E}\left[\mathbf{c}(\mathbf{X}_i) - \boldsymbol{\theta}_0 | S_i = 0 \right] = 0$. This implies $\boldsymbol{\gamma}_0 = -\boldsymbol{\gamma}^{\dag}$. This result is more concretely demonstrated by evaluating 
\[\begin{split} 
\mathbb{E}\left[\boldsymbol{\zeta}_1(S_i, \mathbf{X}_i; \boldsymbol{\gamma}_1, \boldsymbol{\theta}_0)\right] &= \mathbb{E}\left(\mathbb{E}\left\{S_i\exp\left[-\sum_{j = 1}^m c_j(\mathbf{X}_i)\gamma_{1j}\right]\left[\mathbf{c}(\mathbf{X}_i) - \boldsymbol{\theta}_0\right]\middle|\mathbf{X}_i\right\}\right) \\
&= \mathbb{E}\left\{\frac{\exp\left[-\sum_{j = 1}^m c_j(\mathbf{X}_i)(\gamma_{1j} - \gamma^{\dag}_j)\right]}{1 + \exp\left[\sum_{j = 1}^m c_j(\mathbf{X}_i)\gamma^{\dag}_j\right]}\left[\mathbf{c}(\mathbf{X}_i) - \boldsymbol{\theta}_0\right]\right\}.
\end{split}\] If $\boldsymbol{\gamma}_{1} = \boldsymbol{\gamma}^{\dag}$ then $\mathbb{E}\left[\boldsymbol{\zeta}_1(S_i, \mathbf{X}_i; \boldsymbol{\gamma}_1, \boldsymbol{\theta}_0)\right] = \mathbb{E}\left\{[1 - \rho(\mathbf{X}_i)]\left[\mathbf{c}(\mathbf{X}_i) - \boldsymbol{\theta}_0\right]\right\} \propto \mathbb{E}\left[\mathbf{c}(\mathbf{X}_i) - \boldsymbol{\theta}_0\middle|S_i = 0\right] = 0$. Evaluating the estimating equations for $\boldsymbol{\gamma}_{0}$ and $\boldsymbol{\gamma}_{1}$ helps simplify solving the expectations of the estimating equations for $\boldsymbol{\delta}_0$ and  $\boldsymbol{\lambda}_1$, the latter of which can be written as
\begin{equation}\label{xi_1:04}
\begin{split} 
\mathbb{E}\left[\boldsymbol{\xi}_1(S_i, \mathbf{X}_i, Z_i; \boldsymbol{\gamma}_1, \boldsymbol{\delta}_1, \boldsymbol{\theta}_0)\right] 
&= \mathbb{E}\left(\mathbb{E}\left\{S_i Z_i\exp\left[-\sum_{j = 1}^m c_j(\mathbf{X}_i)(\gamma_j + \delta_{1j})\right] \left[\mathbf{c}(\mathbf{X}_i) - \boldsymbol{\theta}_0\right]\right\}\right) \\
&\propto \mathbb{E}\left(\mathbb{E}\left\{Z_i\exp\left[-\sum_{j = 1}^m c_j(\mathbf{X}_i)\delta_{1j}\right] \left[\mathbf{c}(\mathbf{X}_i) - \boldsymbol{\theta}_0\right]\middle| S_i = 0, \mathbf{X}_i\right\}\right) \\
&= \mathbb{E}\left\{\frac{\exp\left[-\sum_{j = 1}^m c_j(\mathbf{X}_i)\delta_{1j}\right]}{1 + \exp\left[-\sum_{j = 1}^m c_j(\mathbf{X}_i)\lambda^{\dag}_{0j}\right]} \left[\mathbf{c}(\mathbf{X}_i) - \boldsymbol{\theta}_0\right]\middle| S_i = 0\right\}.
\end{split}
\end{equation}
The only way for (\ref{xi_1:04}) to evaluate to zero is if \[ \frac{\exp\left[\sum_{j = 1}^m c_j(\mathbf{X}_i)\left(\delta^{\dag}_{1j} - \delta_{1j}\right)\right]}{\exp\left[\sum_{j = 1}^m c_j(\mathbf{X}_i)\delta^{\dag}_{0j}\right] + \exp\left[\sum_{j = 1}^m c_j(\mathbf{X}_i)\delta^{\dag}_{1j}\right]} = r_1 \] where $r_1$ is a constant. Under a similar derivation, in order for $ \mathbb{E}\left[\boldsymbol{\xi}_0(S_i, \mathbf{X}_i, Z_i; \boldsymbol{\gamma}, \boldsymbol{\delta}_0, \boldsymbol{\theta}_0)\right]$ to equal zero, we require \[ \frac{\exp\left[\sum_{j = 1}^m c_j(\mathbf{X}_i)\left(\delta^{\dag}_{0j} - \delta_{0j}\right)\right]}{\exp\left[\sum_{j = 1}^m c_j(\mathbf{X}_i)\delta^{\dag}_{0j}\right] + \exp\left[\sum_{j = 1}^m c_j(\mathbf{X}_i)\delta^{\dag}_{1j}\right]} = r_0 \] where $r_0$ is another constant. Moreover, while operating under these constancy conditions, we know that for $\mathbb{E}\left[\phi(S_i, \mathbf{X}_i, Y_i, Z_i)\right]$ to equal zero, we need $r_1 = r_0 = r$. With a little algebra, we solve the above system of equations to conclude $\delta_{1j} = \delta^{\dag}_{1j}$ and $\delta_{0j} = \delta^{\dag}_{0j}$ which also implies \[ \exp\left[\sum_{j = 1}^m c_j(\mathbf{X}_i)\delta^{\dag}_{0j}\right] + \exp\left[\sum_{j = 1}^m c_j(\mathbf{X}_i)\delta^{\dag}_{1j}\right] = r^{-1}. \] Combining these results, we find that
\[\begin{split}
\mathbb{E}\left[\boldsymbol{\xi}_0(S_i, \mathbf{X}_i, Z_i; \boldsymbol{\gamma}_1, \boldsymbol{\delta}_1, \boldsymbol{\theta}_0)\right] &= r\mathbb{E}\left[\mathbf{c}(\mathbf{X}_i) - \boldsymbol{\theta}_0\middle| S_i = 0\right] = 0, \\
\mathbb{E}\left[\boldsymbol{\xi}_1(S_i, \mathbf{X}_i, Z_i; \boldsymbol{\gamma}_0, \boldsymbol{\delta}_0, \boldsymbol{\theta}_0)\right] &= r\mathbb{E}\left[\mathbf{c}(\mathbf{X}_i) - \boldsymbol{\theta}_0\middle| S_i = 0\right] = 0, \enskip \text{and} \\
\mathbb{E}\left[\phi(S_i, \mathbf{X}_i, Y_i, Z_i; \boldsymbol{\gamma}_0, \boldsymbol{\gamma}_1, \boldsymbol{\delta}_0, \boldsymbol{\delta}_1, \tau_0)\right] &= r\mathbb{E}\left[Y_i(1) - Y_{i}(0) - \tau_0\middle| S_i = 0\right] = 0. \end{split} \] 

Concatenating the parameter values into $\boldsymbol{\nu} = (\boldsymbol{\theta}_0^T, \boldsymbol{\gamma}_0^T,  \boldsymbol{\gamma}_1^T, \boldsymbol{\delta}^T_0, \boldsymbol{\delta}^T_1, \tau_0)^T$ and stacking the estimating equations with \[ \boldsymbol{\psi}(S_i, \mathbf{X}_i, Y_i, Z_i;\boldsymbol{\nu}) \equiv [\boldsymbol{\omega}^T(S_i, \mathbf{X}_i),\boldsymbol{\zeta}^T_0(S_i, \mathbf{X}_i), \boldsymbol{\zeta}^T_1(S_i, \mathbf{X}_i), \boldsymbol{\xi}^T_0(S_i, \mathbf{X}_i, Z_i), \boldsymbol{\xi}^T_1(S_i, \mathbf{X}_i, Z_i), \phi(S_i, \mathbf{X}_i, Y_i, Z_i)], \] we can see that $\tilde{\boldsymbol{\nu}}  = \left(\hat{\boldsymbol{\theta}}_0^T, \tilde{\boldsymbol{\gamma}}_0^T, \tilde{\boldsymbol{\gamma}}_1^T, \tilde{\boldsymbol{\delta}}^T_0, \tilde{\boldsymbol{\delta}}^T_1, \tilde{\tau}_{\text{CAL}}\right)^T$ is an M-estimator for $\boldsymbol{\nu}$. Under mild regularity conditions about M-estimators \citep{tsiatis_semiparametric_2006}, we know that
\begin{equation}\label{influence:04}
\sqrt{n}\left(\tilde{\boldsymbol{\nu}} - \boldsymbol{\nu}\right) = -\mathbb{E}\left[ \frac{\partial\psi(S_i, \mathbf{X}_i, Y_i, Z_i;\boldsymbol{\nu})}{\partial\boldsymbol{\nu}}\right]^{-1} \left[\frac{1}{\sqrt{n}} \sum_{i = 1}^n \psi(S_i, \mathbf{X}_i, Y_i, Z_i;\tilde{\boldsymbol{\nu}})\right] + o_p(1).
\end{equation} 
Equation (\ref{influence:04}) is known as the influence function for $\boldsymbol{\nu}$ and implies $\mathbb{E}\left(\tilde{\boldsymbol{\nu}} - \boldsymbol{\nu}\right) = 0$. By applying the weak law of large numbers, we conclude $\tilde{\boldsymbol{\nu}} \rightarrow_p \boldsymbol{\nu}$ implying $\tilde{\tau}_{\text{CAL}} \rightarrow_p \tau_0$. 

Another result of M-estimation theory shows that, under the weak law of large numbers, \[ \sqrt{n}(\tilde{\boldsymbol{\nu}} - \boldsymbol{\nu}) \rightarrow_d \mathcal{N}(\mathbf{0}_{4m + 1}, \boldsymbol{\Sigma}) \] where \[ \boldsymbol{\Sigma} = \mathbb{E}\left[ \frac{\partial\psi(S_i, \mathbf{X}_i, Y_i, Z_i;\boldsymbol{\nu})}{\partial\boldsymbol{\nu}}\right]^{-1} \mathbb{E}\left[\psi(S_i, \mathbf{X}_i, Y_i, Z_i;\boldsymbol{\nu})^{\bigotimes2}\right] \mathbb{E}\left[ \frac{\partial\psi(S_i, \mathbf{X}_i, Y_i, Z_i; \boldsymbol{\nu})}{\partial\boldsymbol{\nu}}\right]^{-T}. \] Therefore, a robust variance estimator for $\tilde{\boldsymbol{\nu}}$ is \[ \tilde{\boldsymbol{\Sigma}} = \frac{1}{n}\left[ \sum_{i = 1}^n \frac{\partial\psi(S_i, \mathbf{X}_i, Y_i, Z_i;\tilde{\boldsymbol{\nu}})}{\partial\boldsymbol{\nu}}\right]^{-1} \left[\sum_{i = 1}^n \psi(S_i, \mathbf{X}_i, Y_i, Z_i;\tilde{\boldsymbol{\nu}})^{\bigotimes2}\right] \left[ \sum_{i = 1}^n \frac{\partial\psi(S_i, \mathbf{X}_i, Y_i, Z_i; \tilde{\boldsymbol{\nu}})}{\partial\boldsymbol{\nu}}\right]^{-T}.\] The last diagonal element of $\boldsymbol{\Sigma}$ represents the variance for $\tilde{\tau}_{\text{CAL}}$, and is estimated by the last diagonal element of $\tilde{\boldsymbol{\Sigma}}$.

The transportability scenario requires slight modifications to some of the estimating equations in Section \ref{proof:04}. The score equations for $\boldsymbol{\delta}_0$ and $\boldsymbol{\delta}_1$ are replaced by the score equation for $\boldsymbol{\delta}$ defined as
\[\boldsymbol{\xi}(S_i, \mathbf{X}_i, Z_i; \boldsymbol{\gamma}, \boldsymbol{\delta}, \boldsymbol{\theta}_0) \equiv S_i Z_i\exp\left[- \sum_{j = 1}^m c_j(\mathbf{X}_i)(\gamma_j + \delta_{j})\right]\left[\mathbf{c}(\mathbf{X}_i) - \boldsymbol{\theta}_0\right].\]
Following a similar theme, we also rewrite the score equations for $\boldsymbol{\gamma}$, which replaces $\boldsymbol{\gamma}_0$ and $\boldsymbol{\gamma}_1$, as 
\[\boldsymbol{\zeta}(S_i, \mathbf{X}_i; \boldsymbol{\gamma}, \boldsymbol{\theta}_0) \equiv S_i\exp\left[- \sum_{j = 1}^m c_j(\mathbf{X}_i)\gamma_j\right]\left[\mathbf{c}(\mathbf{X}_i) - \boldsymbol{\theta}_0\right]\]
and for $\tau_0$ as
\[\begin{split}
     \phi(S_i, \mathbf{X}_i, Y_i, Z_i; \boldsymbol{\gamma}, \boldsymbol{\delta}, \tau_0) &\equiv S_i Z_i\exp\left[- \sum_{j = 1}^m c_j(\mathbf{X}_i)(\gamma_j + \delta_{j})\right]\left[Y_i(1) - \tau_0\right] \\ 
     &\qquad- S_i(1 - Z_i)\exp\left[- \sum_{j = 1}^m c_j(\mathbf{X}_i)(\gamma_j + \delta_{j})\right]Y_i(0)
\end{split} \]
With these updated estimating equations, the consistency proofs for transportability are nearly the same as the consistency proofs for data-fusion case.

\section{Risk of Total Mortality from Insulin Provision Versus Sensitization in a Veteran Population with Diabetes and CAD}

The second scenario is particularly apt as the last several years has seen the development and approval of a number of new type 2 diabetes medications.\cite{zinman_empagliflozin_2015, marso_semaglutide_2016, marso_liraglutide_2016,  neal_canagliflozin_2017, holman_effects_2017} Valid methods of transporting effect estimates from randomized trials to real world patient populations are essential to maximizing the population health impact of these new medications.

As a counterpoint to the example transporting results within the VA system across two temporally distinct cohorts, we will also apply the methods developed in this paper to transport results from a randomized trial comparing two diabetes treatment strategies, the Bypass Angioplasty Revascularization Investigation 2 Diabetes (BARI 2D) trial, to the aforementioned 2010-2014 cohort of veterans with diabetes.\cite{the_bari_2d_study_group_randomized_2009} The BARI 2D trial attempted to address a second pressing knowledge gap pertaining to type 2 diabetes treatment: whether diabetes patients with coronary artery disease (CAD), the most common underlying cause of mortality for diabetes patients,\cite{rao_kondapally_seshasai_diabetes_2011} benefited from an insulin sensitization or insulin provision strategy for treating their diabetes. Thus, the BARI 2D trial randomized diabetes patients with known CAD to receive either an insulin sensitization strategy (largely treatment with metformin) or an insulin provision strategy (treatment with sulfonylureas and/or insulin). Given the importance of optimal glycemic management in the particularly high-risk population of diabetes patients with CAD, evaluating transportability of results from BARI 2D to a real-world population of VA diabetes patients could provide applicable insight into diabetes population health management within the VA health system. This second demonstration scenario highlights a particularly important transportability application for contemporary diabetes care which has been transformed by randomized trial evidence of efficacy of several new classes of medications with as yet limited real-world data to assess effectiveness.

In addition to transporting the risk difference of total mortality among patients receiving sulfonylurea versus metformin across temporally defined populations within the VA, we also transport estimates from the BARI 2D trial population onto the VA 2010-2014 cohort. This example showcases the fact that our method works for transporting estimates from trial data onto observational data in addition to between observational samples as shown in Section 6 of the main manuscript. In this example we compare insulin sensitization therapy, which consists of treatment with metformin and/or a thiazolidinedione (another class of oral diabetes medications), with insulin provision therapy, which consists of treatment with a sulfonylurea and/or insulin, on the risk of total mortality three years after randomization.

\begin{figure}[ht]
	\centering
	\includegraphics[scale = 0.5]{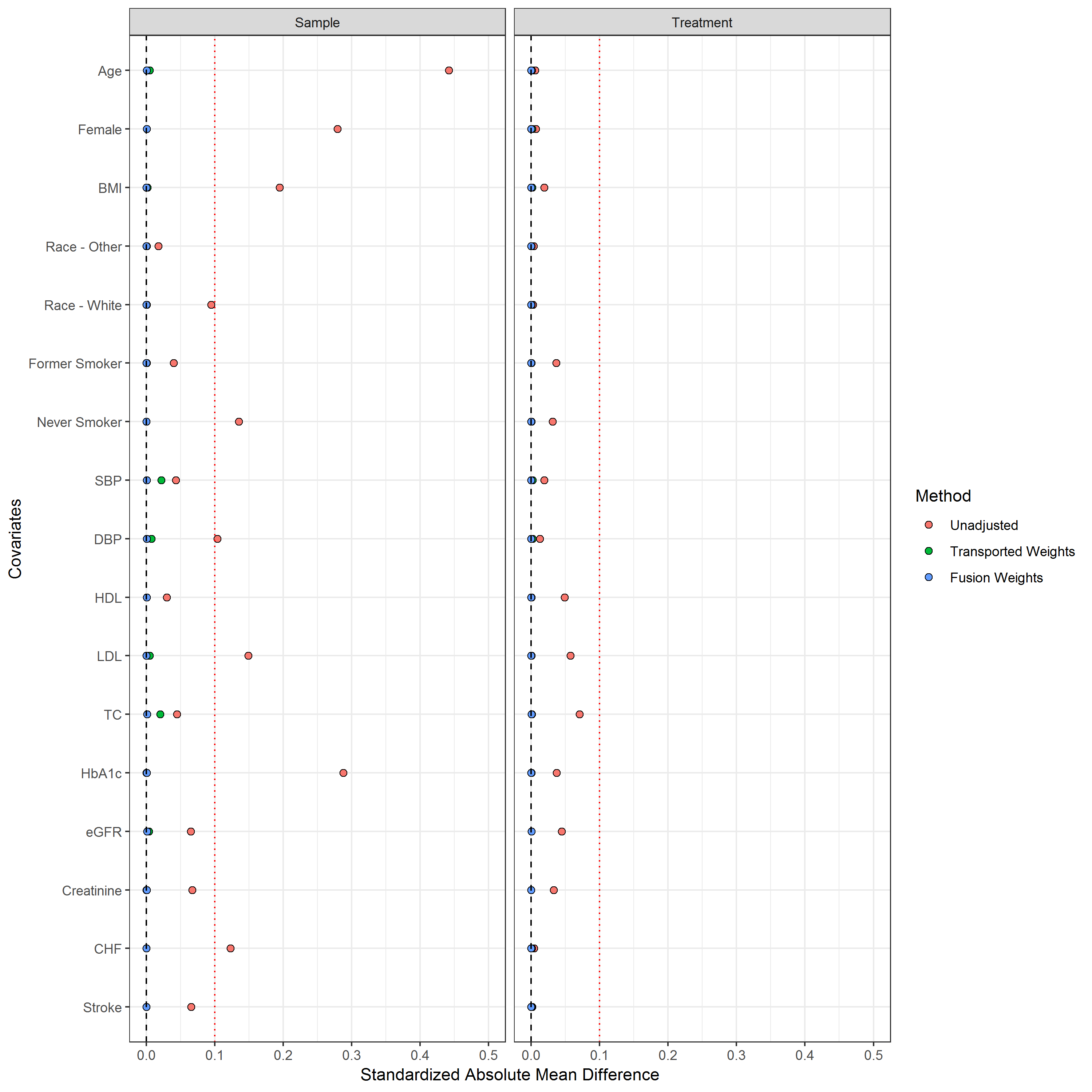}
	\caption{The standardized absolute mean differences between samples (BARI 2D versus 2010-2014 new VA diabetes patients) and treatment groups (insulin sensitization versus insulin provision) using weighting methods discussed in Sections 3.3 and 4. The standardized mean differences between treatment groups are estimated over the patients $\{i:S_i = 1\}$ with the Unadjusted and Transport Weights but for all $i = 1,2,\ldots,n$ with the Fusion Weights. The entire dataset is used to find the differences between samples for all three methods.}\label{cobalt-plot:04}
\end{figure}

The BARI 2D study enrolled 2,368 diabetes patients with untreated coronary artery disease (CAD) into four treatment groups along a 2x2 factorial design.  In addition to examining the effects of glycemic control strategies, the BARI 2D study also tested the effects of delayed versus contemporaneous treatment of CAD. We will ignore this portion of the study and focus solely on glycemic control. We construct a representative cohort from the VA electronic health record as the target sample. This sample included all diabetes patients diagnosed within the calendar years of 2010-2014 with prior history of CAD that received either insulin provision or insulin sensitization therapy after diagnosis with diabetes mellitus (n = 30,393). We note that there is some misalignment with the two samples in this example since patients within the BARI 2D study had a longer duration of diabetes prior to randomization, during which time most patients had some diabetes treatment prior to randomization. If we were to try and balance this variable between samples, we would likely violate Assumption 3. Nevertheless we should still be able to estimate informative risk differences after accounting for the other risk factors attributable to total mortality considered for diabetes patients and after accounting for the covariates with differences between samples in our model. The factors which we balance between the two samples and the treatment groups are found in Figure \ref{cobalt-plot:04}. Here we display the standardized absolute mean differences of the covariates between the treated and controls as well as between the BARI 2D sample and the sample of 2010-2014 VA diabetes patients with CAD, both before and after adjustment using weights for transportability and data-fusion. These covariates are measured in both the BARI 2D trial sample and within the VA cohort. Note that after weighting, these statistics approximately estimate zero by virtue of the primal problems from which they are derived.

Similar to the analysis in Section 6, we find the unadjusted and the CBPS adjusted risk difference estimates in both the VA and BARI 2D cohorts. We then transport the BARI 2D results to the VA cohort using the calibration methods discussed in Section 6. We supplement this estimate by finding the risk difference under the data-fusion setting which combines the responses, treatment, and covariates of the BARI 2D study with the VA cohort. Figure \ref{cobalt-plot:04} offers a visual representation of what happens to a common balance diagnostic (the standardized absolute mean difference) after weighting. Both the crude and adjusted results using data from the BARI 2D study without integrating VA data corroborate what was originally found in the trial analysis - insulin provision has no effect on total mortality compared to insulin sensitization. After three years and adjusting using the CBPS weights, the BARI 2D study saw no change (-2.1\%, 2.2\%) in total mortality between the two treatment groups. However, after weighting the BARI 2D responses to transport the estimated risk difference onto the VA cohort, we observed a 2.4\% (-4.1\%, 8.8\%) increased risk of death among patients receiving insulin provision therapy. For the data-fusion result, we estimate an increase risk in total mortality of 4.2\%  (2.9\%, 5.5\%) three years after randomization. This better aligns with the risk difference estimated using only the VA cohort which found an increased risk difference of 4.2\% (3.0\%, 5.4\%). This result is primarily driven by the much larger size of VA cohort. Notice also that the confidence intervals of the data-fusion estimator are slightly wider than those of the CBPS estimator. This result is interesting from a methodological standpoint as it reflects the added uncertainty inserted by the transported BARI-2D estimates relative to the CBPS estimate.

\end{document}